\documentclass[format=acmlarge]{acmart}
\usepackage{amsmath}
\usepackage{kantlipsum} 
\usepackage{mwe}
\usepackage{adjustbox,multirow,array}
\usepackage{wrapfig}
\usepackage{float}
\usepackage{subcaption}
\usepackage[ruled,linesnumbered]{algorithm2e}

\SetAlFnt{\small}
\SetAlCapFnt{\small}
\SetAlCapNameFnt{\small}
\SetAlCapHSkip{0pt}
\IncMargin{-\parindent}
\newcommand{\eg}{{\it e.g., }}
\newcommand{\etal}{{\it et~al. }}
\newcommand{\ie}{{\it i.e., }}
\newcommand{\name}{{ClusPr}}
\newcommand{\mtx}[1]{\ensuremath{\mathbf{#1}}}
\usepackage{fixltx2e}
\usepackage{flexisym}

\usepackage{mathtools, setspace,lscape,algorithmic,verbatim,color, alltt, amsfonts, wrapfig, boxedminipage,pifont,mathtools }
\usepackage{dblfloatfix}

\usepackage{xcolor}
\usepackage{balance}
\usepackage{fancyhdr, lastpage}
\usepackage{subcaption}

\acmJournal{TOIT}
\acmVolume{9}
\acmNumber{4}
\acmArticle{01}
\acmYear{2018}
\acmMonth{1}
\copyrightyear{2018}
\graphicspath{{./figs/}}
\setcopyright{acmlicensed}

\begin{document}
\title{Privacy-Preserving Clustering of Unstructured Big Data for Cloud-Based Enterprise Search Solutions}

\author{SM Zobaed}
\affiliation{%
	\institution{University of Louisiana at Lafayette}
	\department{School of Computing and Informatics}
	\streetaddress{301 E. Lewis Street}
	\city{Lafayette}
	\state{LA}
	\postcode{70503}
	\country{USA}
}

\author{Mohsen Amini Salehi}
\affiliation{%
	\institution{University of Louisiana at Lafayette}
	\department{School of Computing and Informatics}
	\streetaddress{301 E. Lewis Street}
	\city{Lafayette}
	\state{LA}
	\postcode{70503}
	\country{USA}
}

\begin{abstract}
Cloud-based enterprise search services (\eg Amazon Kendra) are enchanting to big data owners by providing them with convenient search solutions over their enterprise big datasets. However, individuals and businesses that deal with confidential big data (\eg credential documents) are reluctant to fully embrace such services, due to valid concerns about data privacy. Solutions based on client-side encryption have been explored to mitigate privacy concerns. Nonetheless, such solutions hinder data processing, specifically clustering, which is pivotal in dealing with different forms of big data. For instance, clustering is critical to limit the search space and perform real-time search operations on big datasets. To overcome the hindrance in clustering encrypted big data, we propose privacy-preserving clustering schemes for three forms of unstructured encrypted big datasets, namely static, semi-dynamic, and dynamic datasets. To preserve data privacy, the proposed clustering schemes function based on statistical characteristics of the data and determine (A) the suitable number of clusters and (B) appropriate content for each cluster. Experimental results obtained from evaluating the clustering schemes on three different datasets demonstrate between 30\% to 60\% improvement on the clusters' coherency compared to other clustering schemes for encrypted data. Employing the clustering schemes in a privacy-preserving enterprise search system decreases its search time by up to 78\%, while increases the search accuracy by up to 35\%. 

\end{abstract}

%
%


%
%



\thanks{
	We thank the anonymous reviewers of the manuscript. This is a substantially extended version of a paper presented at the Proceedings of the 21st IEEE International Conference on High Performance Computing and Communications (HPCC '19)~\cite{clustcrypt}.
	This research was supported by the Louisiana Board of Regents under grant number LEQSF(2017-20)-RD-B-06 and Perceptive Intelligence, LLC}

\maketitle

\renewcommand{\shortauthors}{Zobaed et al.}

\section{Introduction}\label{sec:intro}
 It is estimated that every day 2.5 Exabytes of both structured and unstructured data are being generated from various sources, such as sensors in weather/traffic/flight systems, organizational documentation/reports, emails, web pages, social media activities (\eg Facebook and Twitter), digital pictures/videos, and transaction records~\cite{exa}. The massive volume of generated data is often referred to as \emph{big data}. It is estimated that more than 95\% of the big data is in unstructured (\emph{document set}) form~\cite{clustcrypt}.
\newline
\indent Cloud providers recently offer scalable and convenient \emph{enterprise search} services (\eg Amazon Kendra \cite{kendra}) to enable searching over their enterprise big document sets (datasets), stored in the cloud. Such services periodically crawl in the dataset and update their index structure that is used for searching. Specifically, a query is searched against the index and the result-set, referencing the relevant documents with respect to the query, is retrieved for the user.
However, using enterprise search services implies outsourcing contents to the cloud that in certain domains, such as law-enforcement and healthcare reports, have raised serious data privacy concerns \cite{S3BD,al2019cybersecurity}, particularly,
after numerous recent data privacy violations \cite{zobaedbig} in the cloud environments. In one incident~\cite{verizonacci}, more than 14 million Verizon customer accounts information were exposed from their cloud repository in 2018. In another incident~\cite{yahooacci}, confidential information of over three billion Yahoo users were exposed. 
\newline
\indent An ideal solution for organizations should enable them to securely store their documents in the cloud while providing real-time enterprise search ability to their authorized users who potentially use thin-clients (\eg hand-held devices and smartphones) with storage and processing constraints. 
Client-side encryption~\cite{S3C,S3BD}, in which documents are encrypted with the user's key before outsourcing to the cloud, is a promising method to achieve the desired data privacy. In this method, only the client has the ability to decrypt the documents, hence, it can potentially provide confidentiality against both internal and external attackers~\cite{kaaniche2014secure,zobaedbig}. However, the limitation of this method is the lack of processing (\eg search and clustering) ability on the encrypted documents. 
\newline
\indent Searchable Encryption systems (\eg~\cite{pham2019survey,cui2016key,zhang2020lightweight,zhang2020se,salehi2017reseed}) have been developed to enable privacy preserving search ability over encrypted data. Such systems predominantly extract keywords (aka tokens) from documents to build an encrypted index, which is  traversed against a search query at the search time to find relevant documents. The problem arises for big datasets where the index structure is prohibitively large, such that traversing it becomes the search bottleneck~\cite{S3BD}. One approach to resolve the bottleneck is topic-based clustering of the index structure, thereby, pruning the search space and limiting it only to the search query context.
\newline
\indent However, clustering encrypted data is challenging because the data semantic is lost, once it is encrypted. The clustering challenge gets further complicated when we consider dynamism that exists in some big datasets. We define a \textit{static dataset} as the one that remains unchanged, once it is uploaded to the cloud. Obsoleted data (\eg HUB5~\cite{da}), archived data (\eg ACCC, RFC~\cite{S3BD}), newsgroup data, book contents are the examples of such datasets. Alternatively, a \textit{dynamic dataset} refers to the ones whose document set grows or shrinks over time, such as criminal records\cite{sa2}, digital libraries \cite{moma21}, and social network feeds \cite{sa3}. 
\newline
\indent Prior research works (\eg~\cite{S3BD,XuCroft, BGoogle}) suggest that statistical characteristics of keywords that are present in the document set (\eg keyword co-occurrences across different documents) can be utilized to cluster the keywords of an encrypted document set. Accordingly, in this research, our \emph{goal} is to develop a topic-based clustering mechanism, called \name, for various types of unstructured big datasets with sensitive contents. In particular, considering the dynamism that potentially exists in certain datasets, we propose two variants of \name~that are tailored for static and dynamic datasets. 
Unlike straightforward \textit{K}-means clustering algorithm \cite{kmeans} where number of clusters and cluster-centers are initialized arbitrarily, we propose two different types of heuristics to cluster static and dynamic datasets. We evaluate the clustering schemes on three different datasets, in terms of the number and coherency of resulting clusters. 
We deploy our clustering methods within a secure cloud-based enterprise search system, called S3BD \cite{S3BD}, and show the advantage of our clustering schemes in improving relevancy and timeliness of the search results.
\newline
\indent In sum, the contributions of this research are as follows:

\begin{itemize}
    \item We propose a method to estimate the suitable number of clusters (\textit{K}) needed to partition a given encrypted dataset (see \cref{subsec:kestimate}).
    \item To enable privacy-preserving clustering in the cloud, we develop three clustering schemes (namely, S-\name, SD-\name, and FD-\name) that can cluster encrypted data across static, semi-dynamic, and fully-dynamic unstructured datasets, respectively (see \cref{subsec:sdcs}, \cref{subsec:fdcs}).
    \item We evaluate and analyze the developed clustering schemes against both encrypted and plain-text clustering approaches with respect to the cluster goodness metrics (see \cref{subsec:evaluationres}).
\end{itemize}
The rest of the paper is organized as follows. In Section \ref{sec-background}, we discuss about background study and related prior works. We explain the overview of our proposed system architecture in Section \ref{sec: propo}. Then, in Section \ref{sec: seca}, we provide a review on the considered threat model and provide security analysis. We discuss about result comparison and performance analysis in Section \ref{sec:evltn}. Finally, Section \ref{sec:conclsn} concludes the paper.

\section{Background and Related work}
\label{sec-background}
\subsection{Fundamental Data Clustering Algorithms}
Once the number of clusters is determined, a dataset is ready to be clustered by utilizing various clustering algorithms~\cite{kmeans,LiuCroft,comjnl/32.3.193, coates2012learning,pelleg2000x}. The tokens are distributed into the cluster with respect to the most similar cluster center. Popular clustering algorithms, such as \textit{K}-means, \textit{X}-means build clusters based on the convergence of center shifting. For a dataset with $n$ datapoints, $K$ clusters, $i$ iterations, and $f$ features, provided that $k \times i \times f \leq n $, the time complexity of \textit{K}-means is $\mathcal{O}(n \times k\times i\times f) \approx \mathcal{O}(n^2)$~\cite{kcomplexity,kcom}. In addition, Elbow method~\cite{elbow}, Silhouette coefficient~\cite{rousseeuw1987silhouettes}, and Bayesian Information Criterion (BIC) scores~\cite{wang2020mixture} are commonly utilized to identify the suitable cluster set. These procedures iteratively build different sets of clusters and nominate a final set based on the minimum loss with respect to the clusters' centers. Adopting any of these procedures increase the time complexity by $K$ times. Hence, the total complexity stands to $\mathcal{O}(Kn^2)$ that is not tractable for big data~\cite{jung2019efficiency, kcomplexity}. To avoid this, we propose two sub-tasks: first, from the index structure, we find the potential centers with $\mathcal{O}(n \times c)$ time complexity, where $c$ denotes the maximum number of centers and we have $c <<n$; Second, we build a cluster-wise token distribution function to assign each token to a proper cluster. Let $q$ denote the number of appropriate centers (and we have $q \leq c$). Then, the overall time complexity for the two sub-tasks is $\mathcal{O}(2n\times c)$.   
\newline
\indent The idea of topic-based clustering has been studied and applied extensively on plain-text datasets. 
Correspondingly, Xu and Croft \cite{XuCroft} proposed to build clusters on a homogeneous index (\ie all of the terms share a nearly similar topic) that improved the effectiveness of a search system compared to standard distributed information retrieval systems. 
The authors used \textit{K}-means clustering algorithm, and to assign components among the clusters, they used KL-divergence \cite{kullback1951information}. Then, utilizing the maximum likelihood estimation theory, their proposed method determines the highest relevant cluster based on the incoming search query. However, the overall process is computationally intensive that impacts the real-time search over big data.
\newline
\indent Mary and Kumar~\cite{mary2012density} addressed the challenges of clustering dynamic data (\eg Twitter and streaming data). They utilized Density-Based Spatial Clustering of Applications with Noise (DBSCAN) algorithm~\cite{tran2013revised}. However, DBSCAN falls short in clustering datasets containing different densities, which is common in big data~\cite{mddb}. Alternatively, \name~clusters datasets irrespective to their density. 
\subsection{Privacy-Preserving Data Clustering Schemes}
A large body of research has been undertaken to enable processing of the encrypted data (ciphertext). Zhou~\etal~proposed a linear transformation-based solution for matching queries against encrypted data while ensuring data privacy on the cloud without any intervention of the data owner~\cite{zhou2017efficient}. However, linear transformation methods support secure \textit{K}-nearest neighbor (KNN)-based query matching approaches but not the clustering. This is because clustering is not invariant to linearly transformed data. 
The optimal linear transformation has a prerequisite of knowing the true cluster means, which is not possible to obtain before generating the cluster~\cite{tarpey2007linear}.
In addition, we assume that the data are tokenized and encrypted before transferring to the cloud. Therefore, unlike~\cite{zhou2017efficient}, where the entirety of encrypted data is queried using time-consuming cryptographic calculations, we use the statistical properties of the data without revealing any meaningful part of it to the cloud. Sun \etal  proposed a searchable encryption method by forming a tree index structure that operates based on the cosine similarity and $TF \times IDF$~\cite{sun2013privacy,sun2014privacy} measures. However, the solution is not scalable for big data, because the search index can become large to the extent that it impacts timeliness of the search operation. We believe that our proposed clustering approach can be a complement to \cite{sun2013privacy,sun2014privacy} where the central index is partitioned topically into multiple small size index structures that can improve the search time and efficiency.
\newline
\indent Homomorphic encryption has become a popular method to perform computation over the encrypted data. Several variations of the homomorphic encryption such as
fully or partially Homomorphic encryption~\cite{homomorphic:modern,xing2017mutual} have been proposed to enable privacy-preserving data processing on the cloud.
Zhu~\etal~\cite{zhu2020privacy}
proposed a secure aggregation and division
protocol based on homomorphic encryption to securely compute
clusters without tampering with the privacy of individual peers in a peer-to-peer
system. However, their clustering technique does not consider data dynamism. Pang and Wang proposed a homomorphic scheme that provides security to outsourced data uploaded from multiple parties in a twin-cloud system~\cite{pang2020privacy} that is assumed to be a semi-honest environment, whereas, we assume cloud to be untrusted in terms of storing/processing sensitive data~\cite{li2018privacy}.
Wang~\etal proposed \textit{HK}-Means++ that combines~\textit{K}-Means clustering with finding the suitable cluster numbers~\cite{wang2019research}. In addition, the work leverages homomorphic encryption scheme to solve the encrypted data manipulation, distance, and convergence calculation. Although our
work is comparable to \textit{HK}-Means++, it can only cluster static datasets. Moreover, the experiments were performed only on one dataset and it is not clear how the method performs on other datasets.
\newline
\indent We note that the current implementations of the homomorphic encryption technique imply a high computational overhead~\cite{homomorphic:slow} which affects the real-time response of a search system, particularly, for big datasets~\cite{zhu2020privacy}.    
\newline
\indent Vaidya and Clifton~\cite{privvad} proposed a solution to cluster encrypted datasets in which different data attributes are stored in distinct storage systems. Then, the clustering was carried out in each one of the data storage systems individually. However, this solution is time consuming and cannot serve the real-time constraint we consider in this work. 
\subsection{Estimating Number of Clusters}
A prerequisite to clustering is to estimate the suitable number of clusters \textit{K}. However, identifying \textit{K} is an NP-hard problem~\cite{Paparrizos}, hence, a large body of research has been undertaken to provide heuristic methods in which \textit{K} is approximated ~\cite{Can1990,clustcrypt,fu2019estimating,coates2012learning}.  
Established clustering methods, such as Silhouette~\cite{rousseeuw1987silhouettes}, Gap Statistic~\cite{tibshirani2001estimating}, Stability Selection~\cite{lange2004stability}, Consensus~\cite{consensus}, and Progeny~\cite{hu2015progeny}, 
generate a series of various clustering sets upon considering various number of clusters. Later, they select the optimal clustering set from the series. The common drawback of these methods is the computational overhead, because the clustering has to be performed multiple times with respect to the different number of clusters.
\newline
\indent Pelleg and Moore~\cite{pelleg2000x} proposed a regularization framework for approximating \textit{K} upon utilizing \textit{X}-means clustering. The method is a modified version of \textit{K}-means clustering that improves cluster distributions by frequently attempting subdivision until it satisfies a predefined stopping criterion. However, lack of prior knowledge about the dataset makes it difficult to apply the associated stopping criterion~\cite{comjnl/32.3.193}. 
In \textit{X}-means method, an optimization function is used to choose the smallest number of clusters with the maximum possible amount of variation within the dataset. The algorithm starts with \textit{K}=1 and increment it until the variation reaches a plateau. This starting point is considered as the optimal \textit{K} for the \textit{K}-means method~\cite{coates2012learning}.
Although the clustering methods are dataset independent, estimation of an appropriate~\textit{K} is generally dataset-specific. \textit{K} should be higher in sparse data and lesser in comparatively dense data. Likewise, we propose to estimate appropriate $K$ with respect to the content of the considered dataset.
\begin{table}[h]
\centering
\resizebox{\linewidth}{!}{
\begin{tabular}{|l|c|l|l|c|c|c|c|}
\hline
\textbf{Research Works}         & \textbf{\begin{tabular}[c]{@{}c@{}}Estimating\\ \#Clusters\end{tabular}} & \textbf{\begin{tabular}[c]{@{}c@{}}Encryption\\ Approach\end{tabular}} & \textbf{\begin{tabular}[c]{@{}c@{}}Cloud's\\ Trustworthiness\end{tabular}} & \textbf{\begin{tabular}[c]{@{}c@{}}Using Edge \\Computing\end{tabular}} & \textbf{\begin{tabular}[c]{@{}c@{}}Real-time \\ Support\end{tabular}} & \textbf{\begin{tabular}[c]{@{}c@{}}Dynamic Data \\ Clustering\end{tabular}} & \textbf{\begin{tabular}[c]{@{}c@{}}Multiple Data\\ Owners\end{tabular}} \\ \hline
 Wang \etal~\cite{wang2019research}      & No                    & Homomorphic         & Semi-honest                                                                  & No                                                                      & No                                                                                       & No                                                                          & No                                                                      \\ \hline
Valdiya \& Clifton~\cite{privvad}      & No                    & Homomorphic         & Semi-honest                                                                  & No                                                                      & Yes                                                                                 & No                                                                          & Yes                                                                     \\ \hline
Pang \& Wang~\cite{pang2020privacy}     & No                    & Homomorphic         & Semi-honest                                                                  & No                                                                      & Yes                                                                                 & No                                                                          & Yes                                                                     \\ \hline
 Sun \etal~\cite{sun2013privacy,sun2014privacy} & No                    & User-side           & Honest-but-curious                                                           & No                                                                      & Yes                                                                                 & No                                                                          & No                                                                      \\ \hline
 Zhu \etal~\cite{zhou2017efficient}                  & No                    & Homomorphic         & Honest                                                                       & No                                                                      & No                                                                                       & No                                                                          & Yes                                                                     \\ \hline
 Woodworth \etal~\cite{S3BD}                 & No                    & User-side           & Honest-but-curious                                                           & No                                                                      & Yes                                                                                      & No                                                                          & Yes                                                                     \\ \hline
ClusPr (proposed)              & Yes                   & User-side           & Honest-but-curious                                                           & Yes                                                                     & Yes                                                                                      & Yes                                                                         & Yes                                                                     \\ \hline
\end{tabular}
}
\caption{Summary of the existing privacy-preserving clustering approaches and positioning our proposed work (ClusPr) with respect to them.}
\label{tab:summary}
\end{table}
\subsection{Positioning of the Proposed Work}
Our proposed work is motivated from Woodworth \etal method for topic-based clustering on encrypted keywords over the central index using \textit{K}-means method \cite{S3BD}. The cluster-wise token distribution function was determined based on the statistical data of each encrypted keyword or token. The authors use a predefined \textit{K} value. Such \textit{K} value is inefficient, because the appropriate number of clusters could be varied based on the dataset characteristics. Moreover, as the authors only considered static/unchanged data, the proposed scheme is not capable of processing dynamic data. On the other hand, \name~provides a heuristic to approximate the suitable number of clusters and then, clustering the data while maintaining the data privacy on the cloud. For a dynamic dataset, where documents are added or removed over time, because of the re-clustering operation, clusters are shrunk or expanded to reflect the dynamism of the dataset.  
\newline
\indent Table~\ref{tab:summary} summarizes the notable related studies in the literature and positions the contribution of this paper with respect to them.  
\begin{figure*}[htp]
  \centering
  \includegraphics[width=0.9\textwidth]{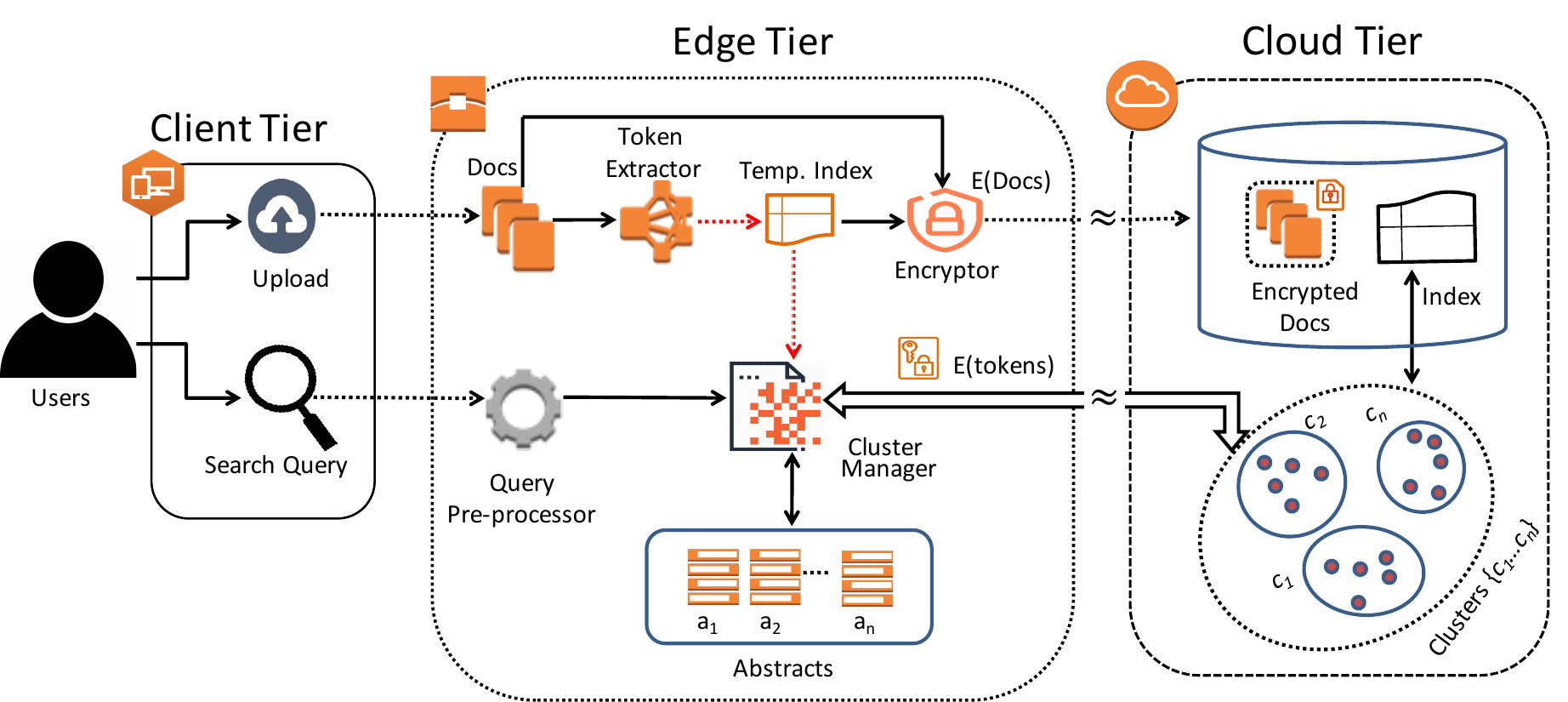}
  \caption{Overview of the context where \name~is deployed in a three-tier architecture (of client, edge, and cloud) to facilitate a secure cloud-based search service. The edge tier is assumed to be on the user premises and trusted. It is used to ease the computational overheads imposed by privacy and clustering related processes.}
  \label{sd}
\end{figure*} 
\section{Three-Tier Architecture of \name} \label{sec: propo}
Figure~\ref{sd} presents an architectural overview of the context where \name~is developed. The architecture represents applying \name~for S3BD, a cloud-based secure semantic search system that requires clustering over encrypted data~\cite{S3BD}. The architecture represents a three-tier system based on a client device, edge system, and the central cloud system. The edge tier resides on the user's premises (hence, is considered trusted) to relieve the client tier from processing computationally intensive tasks. This is particularly important for non-static (\ie semi-, fully-dynamic) datasets where documents have to be processed as they are uploaded to the cloud tier over time. 

In the specific context of S3BD, upon uploading a document by the user, the document is passed through \emph{Token Extractor} on the edge tiers to retrieve the keywords (aka tokens) semantically representing the document. For dynamic datasets, a temporary index structure is used to store the extracted tokens representing the occurrences of each new token in different documents. Next, the document is encrypted by the user's key and is securely stored on the cloud repository. 
Next, a \emph{Temporary Index} structure is formed based on the extracted tokens of the documents in question before encrypting and uploading them to the cloud. The Temporary Index structure shows the tokens, their frequency, and their appearances across the uploaded batch. Tokens of the Temporary Index are encrypted by the \emph{Encryptor} using the user's key. By encrypting documents as well as the extracted tokens, Encryptor preserves the data privacy on the cloud. Note that, although we can technically use homomorphic encryption to maintain the statistical properties (frequency and co-appearances), for efficiency reasons, in the current implementation, we keep the properties unencrypted. We assume that such properties do not reveal meaningful information about the data. In fact, in~\cite{wang2019research}, \textit{K}-means clustering was used over homomorphically encrypted big data and showed that the time overhead of clustering can be prohibitively expensive.  In the next step, the Temporary Index is fed to the \emph{Cluster Manager} to make the suitable clustering decision on the cloud. Cluster Manager may decide to keep the existing clusters and only update them by the entries of the Temporary Index. Alternatively, upon observing a major update in the Temporary Index, the Cluster Manager decides to exhaustively re-cluster all of the tokens. Though a few of the aforementioned prior works can cluster encrypted data, they fall-short in clustering dynamic datasets, whereas, \name~can cluster both static and dynamic data while ensuring privacy. We explain the updating and re-clustering procedures \name~ in Section~\ref{sec:dynamic}. Cluster Manager is also in charge of generating and maintaining \emph{Abstracts}. Each abstract $a_i$ is a sampled summary of a corresponding cluster $C_i$ on the cloud tier \cite{sahan}. Abstracts are to prune the search operation and navigate the search only to clusters that are topically-related to the query. Further details about Abstracts are described in Section \ref{cluster-abs}.  

For static datasets, the architecture is streamlined such that the extracted tokens are encrypted and directly fed into the Index structure on the cloud tier. Once the data uploading procedure is completed, the cloud tier initiates the clustering procedure. As there is no re-clustering procedures defined for static clusters, the Cluster Manager is only in charge of generating and maintaining the abstracts~\cite{sahan}. It is noteworthy that, in the architecture of Figure~\ref{sd}, the dashed arrows located in the edge tier are to highlight the differences for dynamic datasets.

In the implementation, we chose to use RSA~\cite{RSA}, which is a widely-adopted and highly secure deterministic encryption technique, to encrypt the documents and extract their tokens before uploading them to the cloud. Other techniques, such as AES with repeating initialization vector (\eg AES-SIV), can be used for encryption as well. It is noteworthy that probabilistic encryption techniques \cite{kumar2018fully}, such as AES-CBC/GSM \cite{AESCBC}, cannot be used, because they generate different ciphers for the same token.

The index has a key-value structure, where each key is a token and its corresponding value is a set of pairs, each one representing the document that includes the token and the frequency of that particular token. Note that both the keys and document names are encrypted. 
For this reason, in the current implementation, we chose to maintain frequency values in the plain-text format. Further details of the proposed static and dynamic data clustering schemes are presented in Section~\ref{sec:static} and~\ref{sec:dynamic} respectively. 

Cluster $C_i$ on the cloud tier includes a group of index entries that are topically similar. As the indexed tokens are encrypted and do not carry any semantic, topic-based clustering of them is a challenging task. We hypothesize the tokens that are semantically similar tend to appear in the same documents. Accordingly, clustering is performed based on the relative frequency of tokens' co-occurrences across all the documents. The union of \textit{K} topic-based clusters ($C_1...C_k$) is equivalent to the index.  
Upon issuing a search query by the user, the abstracts with the highest similarity to the search query are identified. Then, only the clusters associated with the abstracts are searched. 
\section{Privacy-Preserving Clustering Scheme for Static Big Datasets (S-\name)}
\label{sec:static}
In this part, first (in Section \ref{subsec:kestimate}), we elaborate on how to estimate the appropriate number of clusters that should be formed to represent a static big dataset. Second, in Section \ref{3.3}, we provide an algorithm to form the center of each cluster. Then, in Section \ref{cluster-wise}, we explain methods to distribute the indexed terms across clusters. Finally, in Section \ref{cluster-abs}, we describe the way pruning is achieved, \ie the method that navigates a search query to relevant cluster(s).  
 
\subsection{Estimating the Number of Clusters for Static Big Datasets}
\label{subsec:kestimate}

Depending on the characteristics of a dataset and distribution of tokens in its documents, the appropriate number of clusters ($K$) can vary significantly. However, optimally determining $K$ directly impacts the accuracy of topic-based clustering and, subsequently, the efficiency of the system (\eg search application) that uses the clusters. Encrypted tokens and their metadata, including documents they appear in and their frequency, are the only available parameters to estimate $K$. The tokens and their metadata are generated by
a keyword extractor that retrieves \textit{n} single or multi-phrase tokens from each document. We assume that all documents are treated equally and the value of \textit{n} is the same across all documents in a given static dataset.

Estimating $K$ for the static dataset is performed based on the following two steps: (1) building Token-Document Frequency Matrix; and (2) constructing Normalized Matrix.

\paragraph{\textbf{Step-1: Building Token-Document Frequency Matrix.}}
To be able to follow the scheme, we consider an example using \textit{five} tokens and \textit{six} documents in Table \ref{ini mat}. We initialize a token-document matrix \mtx A from the index structure. In the matrix, each row represents a token and each column represents a document. Although our approach does not deal with plain-text tokens, just for further readability, in the Table \ref{ini mat}, we redundantly show the plain-text tokens (in ``Word" column) along with their encrypted forms (in ``Hash" column).
Each entry $a_{i,j}$ of matrix \mtx A represents the frequency of $i^{th}$ token in $j^{th}$ document (denoted as $f(i,j)$).  
\begin{table}[h]
	\centering
	\caption{Token-Document Frequency Matrix \mtx A, built based on the index structure}
	\label{ini mat}
	\begin{tabular}{|l|l|c|c|c|c|c|c|}
		\hline
		\textbf{Word}       & \textbf{Hash}                                      & \multicolumn{1}{l|}{\textbf{d\textsubscript{1}}} & \multicolumn{1}{l|}{\textbf{d\textsubscript{2}}} & \multicolumn{1}{l|}{\textbf{d\textsubscript{3}}} & \multicolumn{1}{l|}{\textbf{d\textsubscript{4}}} & \multicolumn{1}{l|}{\textbf{d\textsubscript{5}}} & \multicolumn{1}{l|}{\textbf{d\textsubscript{6}}} \\ \hline
		
		\textcolor {gray}{Book}  & \textbf{Uh5W}                                   & 30                               & 0                                & 23                               & 4                                & 40                               & 0                                \\ \hline
		\textcolor {gray}{Solve} & \textbf{/Vdn}                                    & 5                                & 0                                & 0                                & 60                               & 34                               & 0                                \\ \hline
		\textcolor {gray}{Traffic}    & \textbf{oR1r}                                  & 0                                & 23                               & 0                                & 30                               & 0                                & 0                                \\ \hline
		\textcolor {gray}{Net}    & \textbf{vJHZ} & 52                               & 49                               & 0                                & 23                               & 0                                & 26                               \\ \hline
		\textcolor {gray}{Enter}    & \textbf{tH7c}                                  & 0                                & 45                               & 68                               & 0                                & 3                                & 5                                \\ \hline
		
	\end{tabular}
\end{table}

For a big dataset, the matrix size can be prohibitively large and sparse. To avoid this, we trim the matrix to include only the tokens that are influential in building clusters. We define \textit{document co-occurrences} as the number of documents containing a particular token. Then, to build the token-document frequency matrix \mtx A, we only take into account tokens whose document co-occurrences are either greater than or equal to the mean value of the document co-occurrences across the whole dataset. 

\paragraph{\noindent \textbf{Step-2: Constructing Normalized Matrix.}}
To make the relationship among tokens and documents quantifiable and comparable, we need to normalize the token-document frequency matrix. Considering that $a_{i,j}$ represents the strength of association between token $t_i$ and document $d_j$, the maximum value in column $j$ of the token-document frequency matrix represents the token with the highest association with document $d_j$. 
Hence, for normalization, we divide the value of each entry of \mtx A to the highest value in the corresponding column of the matrix and the result is stored in a new matrix, called matrix \mtx N. The value for each entry $n_{i,j}$ is formally calculated based on Equation \ref{fe2}.
\begin{equation}
\label{fe2}
n_{i,j}=\frac{a_{i,j}}{\displaystyle\max_{\forall i} a_{i,j}} 
\end{equation}
\paragraph{\noindent \textbf{Step-3: Building Probabilistic Matrices \mtx R and \mtx S}}
The goal, in this step, is to calculate the topic similarity among encrypted tokens. 
For that purpose, we need to calculate the probability that topic of a token shares similarity with other tokens. We hypothesize that tokens that co-occur across documents are likely to share the same topic. Besides, the magnitude of similarity between two tokens could be influenced by the tokens' distribution across the dataset. For instance, specific terms 
appear only in a few documents and are not widely distributed throughout the dataset. 
Such sparsely distributed tokens have low co-occurrences with other tokens which increases the diversity of topics in a dataset and potentially raises the required number of clusters ($K$). 
We leverage the normalized matrix (\mtx N) to perform a two-phase probability calculation that
 yields a matrix (denoted as \mtx Q) representing token-to-token topic similarity. 
In the first phase, we calculate the \emph{importance} of each token to each document. The importance of token $t_i$, in document $d_j$, denoted as $\tau_{i,j}$, is defined based on Equation~\ref{eq:cont}.
 \begin{equation}\label{eq:cont}
\tau_{i,j} = \frac{n_{i,j}}{\displaystyle\sum_{\forall k} n_{i,k}}   
\end{equation}

Considering Equation~\ref{eq:cont} and matrix \mtx N, we generate matrix \mtx R whose entries represent the importance of each token across all documents. In fact, each entry $r_{i,j}$ of \mtx R represents the probability of choosing a document $d_j$, having token $t_i$. That is, $r_{i,j}=\mathbb P(t_i,d_j)$. 

In the second phase, we calculate the importance of each document to each token. The importance of document $d_j$ for term $t_i$, denoted by $\delta_{j,i}$ and is defined 
based on Equation~\ref{frt}. 
\begin{equation}
\label{frt}
\delta_{j,i} = \frac{n_{j,i}}{\displaystyle\sum_{\forall q} n_{q,i}}
\end{equation}

Considering each $\delta_{j,i}$  
and \mtx N, we generate \mtx S whose entries represent the importance of each document with respect to each token. In fact, each entry $s_{i,j}$ represents the probability of choosing $t_i$ from $d_j$ (\ie we have $s_{i,j}=\mathbb P(d_j,t_i)$). 
\paragraph{\noindent\textbf{Step 4- Constructing Matrix \mtx Q to Determine the Number of Clusters}}

Recall that \mtx R is a token-to-document matrix and \mtx S is a document-to-token matrix. To identify the similarity among the encrypted tokens, we multiply \mtx R and \mtx S. As the number of columns and rows of \mtx R and \mtx S are equal, it is possible to multiply matrix \mtx R with \mtx S. The resultant matrix, denoted as \mtx Q, is a token-to-token matrix and serves as the base to determine the number of required clusters. Each entry $q_{i,j}$ denotes the topic similarity between token $i$ and $j$. More specifically, $q_{i,j}$ indicates the magnitude to which token $i$ shares similar topic with token $j$ for $i\neq j$ and is calculated as $q_{i,j}= \displaystyle\sum_{\forall i,j}r_{i,j}\cdotp s_{j,i}$. 
Table \ref{c mat} shows matrix \mtx Q for the example we discuss in this section.

\begin{table}[H]
	\centering
	
	\caption{Cluster decision matrix \mtx Q is built based on the multiplication of \mtx R and \mtx S matrices}
	\label{c mat}
	\begin{tabular}{|l|c|c|c|c|c|}
		
		\hline
		\textbf{\textcolor{gray}{Word}-Hash}                                                                                       & \multicolumn{1}{l|}{\textbf{\begin{tabular}[c]{@{}l@{}}\textcolor {gray}{Book}\\ Uh5W\end{tabular}}} & \multicolumn{1}{l|}{\textbf{\begin{tabular}[c]{@{}l@{}} \textcolor {gray}{Solve}\\ /Vdn\end{tabular}}} & \multicolumn{1}{l|}{\textbf{\begin{tabular}[c]{@{}l@{}} \textcolor {gray}{Traffic}\\ oRir\end{tabular}}} & \multicolumn{1}{l|}{\textbf{\begin{tabular}[c]{@{}l@{}} \textcolor {gray}{Net}\\ vJHZ\end{tabular}}} & \multicolumn{1}{l|}{\textbf{\begin{tabular}[c]{@{}l@{}} \textcolor {gray}{Enter}\\  tH7c\end{tabular}}} \\ \hline
		\textbf{\begin{tabular}[c]{@{}l@{}}\scriptsize{\textcolor {gray}{Book}- Uh5W}\end{tabular}}                                       & 0.39                                                                                  & 0.25                                                                                  & 0.01                                                                                     & 0.18                                                                                                                       & 0.09                                                                                     \\ \hline
		\textbf{\begin{tabular}[c]{@{}l@{}}\scriptsize{\textcolor {gray}{Solve}- /Vdn}\end{tabular}}                                      & 0.26                                                                                  & 0.45                                                                                  & 0.12                                                                                     & 0.12                                                                                                                       & 0.02                                                                                     \\ \hline
		\textbf{\begin{tabular}[c]{@{}l@{}}\scriptsize{\textcolor {gray}{Traffic}-\ oRir}\end{tabular}}                                  & 0.02                                                                                  & 0.26                                                                                  & 0.21                                                                                     & 0.33                                                                                                                       & 0.18                                                                                     \\ \hline
		\textbf{\begin{tabular}[c]{@{}l@{}}\scriptsize{\textcolor {gray}{Net}- vJHZ}\end{tabular}} & 0.10                                                                                  & 0.07                                                                                  & 0.08                                                                                     & 0.58                                                                                                                       & 0.15                                                                                     \\ \hline
		\textbf{\begin{tabular}[c]{@{}l@{}}\scriptsize{\textcolor {gray}{Enter}-  tH7c}\end{tabular}}                                  & 0.09                                                                                  & 0.01                                                                                  & 0.08                                                                                     & 0.28                                                                                                                       & 0.37                                                                                     \\ \hline
	\end{tabular}
\end{table}


Diagonal entries of \mtx Q signify the topic similarity of each token with itself and dissimilarity (\ie separation) from other topics. More specifically, the value of $q_{i,i}$ indicates the magnitude that term $t_i$ does not share its topic with other terms. Therefore, we define diagonal entries ($q_{i,i}$) as \textit{separation factor}, because for each token, it represents the token's tendency to stay separate from other topics. As such, summation of the separation factors can approximate the number of clusters ($K$) needed to partition topics of a dataset. Let $m$ denote the total number of tokens in \mtx Q. Then, Equation~\ref{eq:trace} is used to approximate $K$ for a given dataset. We use the ceiling function to make $K$ an integer value.

\begin{equation}
k = \lceil\sum_{i=1}^{m} q_{i,i}\rceil
\label{eq:trace}
\end{equation}

Correctness of $K$ is verified using a hypothesis that states $K$ for a set should be higher if individual elements of the set are dissimilar, otherwise $K$ should be low~\cite{Can1990,cutting2017scatter}. Equation~\ref{eq:trace} is the core of approximating $K$. According to this equation, the maximum $K$ value can reach to $M$, when the documents are highly distinct and each individual token of the documents represents a unique topic, otherwise it is lower than $M$. 
Hence, our approach conforms with the clustering hypothesis.

\subsection{Center Selection}
\label{3.3}

In \textit{K}-means clustering, generally, the clusters' centers are arbitrarily chosen~\cite{aggarwal2019performance, LiuCroft}. Then, based on a distance measure function (\eg Euclidean distance~\cite{aggarwal2019performance} or semantic graph \cite{LiuCroft}), dataset elements are distributed into the clusters. \textit{K}-means operates based on iteratively shifting clusters' centers until it converges. However, we realized that the extremely large number of tokens make the iterative center shifting step (and therefore \textit{K}-means clustering) prohibitively time-consuming for big data~\cite{aggarwal2001surprising}. Accordingly, in this part, we are to propose a big-data-friendly method to cluster encrypted tokens. 

The key to our clustering method is to dismiss the iterative center shifting step. This change entails initial clusters' centers not to be chosen arbitrarily, instead, they have to be chosen proactively so that they cover various topics of the dataset. For that purpose, a na\"{i}ve method can choose the top \textit{K} tokens that have the highest number of associated documents. Although this approach chooses important (highly associated) tokens, it ends up selecting centers that have a high topical overlap. We propose to choose tokens that not only have high document association but also cover diverse topics exist in the dataset. 

We define \emph{centrality} of a token $i$, denoted $\Phi_i$, as a measure to represent a topic and relatedness to other tokens of the same topic. Assume that tokens are sorted in a descending manner, based on the degree of document association. Let $U$ represent the union of documents associated to the currently chosen centers. Also, for token $i$, let $A_i$ represent the set of documents associated to $i$. Then, \emph{uniqueness}~\cite{S3BD} of token $i$, denoted $\omega_i$, is defined as the ratio of the number of documents associated to $i$ but not present in $U$ (\ie $|A_i-U|$) to the number of documents associated to $i$ and are present in $U$ (\ie $|A_i\cap U|$). Uniqueness indicates the potential of a token to represent a topic that has not been identified by other tokens already chosen as centers. Particularly, tokens with uniqueness value greater than $1$ have high association to documents that are not covered by the currently chosen centers, hence, can be chosen as new centers.

Recall that each entry $q_{i,j}$ of matrix \mtx Q represents the topic similarity between tokens $i$ and $j$. Besides, diagonal entry $q_{i,i}$ measures separation of token $i$ from others. Therefore, the total similarity token $i$ shares with others can be obtained by $\Sigma_{\forall j | j \neq i}q_{i,j}$. Note that for token $i$, we have $\Sigma_{\forall j}q_{i,j}=1$, hence, the total similarity for token $i$ is equal to $1-q_{i,i}$. Centrality of a token is measured by the uniqueness of the token, the magnitude of similarity the token shares with others, and the magnitude of it being isolated. That is, for token $i$, centrality is defined as: $\Phi_i=\omega_{i}\times q_{i,i}\times (1-q_{i,i})$.

\begin{algorithm}
	\SetAlgoLined\DontPrintSemicolon
	\SetKwInOut{Input}{Input}
	\SetKwInOut{Output}{Output}
	\SetKwFunction{algo}{algo}
	\SetKwFunction{proc}{Procedure}{}{}
	\SetKwFunction{main}{\textbf{ChooseCenter}}
	\SetKwFunction{quant}{\textbf{CalculateUniqueness}}
	\SetKwFunction{docassoc}{\textbf{CalculateDocumentAssoc}}
	\Input{$K$, \mtx C matrix, and $Index$ (with tokens sorted descendingly based on the degree of document association)}
	\Output{ $centers$ set that includes at most $K$ center tokens}
	
	\SetKwBlock{Function}{Function \texttt{ Choose Center($k, \mtx Q, Index$):}}{end}
	
	\Function{
		$centers \gets \emptyset$ \;
		$U \gets \emptyset$ \; 
		$\Theta \gets \{(\emptyset,\emptyset)\}$ //\small{Pairs of tokens and centrality values} \;
		\ForEach{token $i \in Index$} {
			
			$\omega_i \gets \quant(i, U)$\;
			\If {$\omega_i >1$} {	
			    $A_i \gets \docassoc(i,Index)$\;
				$U \gets U \cup A_i$ \; 
				$\Phi_i \gets (\omega_i \times q_{i,i} \times (1-q_{i,i})) $\;
				Add pair ($i,\Phi_i$) to max-heap $\Theta$ based on $\Phi_i$ \; 
			}
			
		}
		
		$centers \gets$ \small{Extract $K$ max pairs from $\Theta$ heap  } \;
		\Return{centers} \;	
		
	}
	\caption{Pseudo-code to determine clusters' centers}
	\label{alg:centr}
	
\end{algorithm}
\vspace{-1pt}
Algorithm \ref{alg:centr} shows the high-level pseudo-code to select maximum of $K$ centers from the set of indexed tokens of a dataset. In addition to $K$, the algorithm receives the central index and the \mtx Q as inputs. The algorithm returns a set of at most $K$ center tokens, denoted $centers$, as output. In the beginning, the output set is initialized to null. $U$ represents the set of documents covered with the chosen centers. A heap structure, denoted $\Theta$, is used to store a pair for each token and its centrality value. 
For each token $i$, the uniqueness and centrality values are calculated (Steps $5 - 13$) and the corresponding pair is inserted to the heap. Note that tokens with uniqueness lower than one do not have the potential to serve as a cluster center. In the next step, we select at most $K$ center tokens that have the highest centrality values.  

\subsection{Distributing Encrypted Tokens Across Clusters}
\label{cluster-wise}

Once $K$ tokens are nominated as cluster centers, the remaining tokens of the index 
are distributed across the clusters with respect to their \emph{relatedness} (a.k.a. distance) with the center tokens. 

\begin{figure} 
 	\centering
 	\includegraphics[width=.7\linewidth]{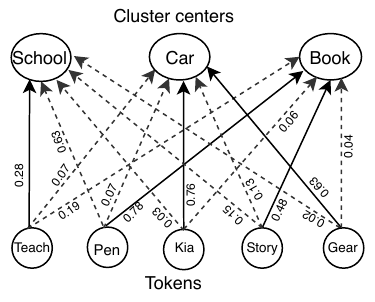}
 	\caption{\small{A bipartite graph representing the relatedness among centers and remaining tokens. The weight of each edge represents the relatedness of a token and a center. Solid lines show centers that offer the maximum relatedness for a token.}}
 	\label{fig: bipartite}
 \end{figure}

Because there is no intersection between the non-center tokens and members of the $centers$ set, we can model the token distribution across the clusters as a weighted bipartite graph where the weight of each edge represents the relatedness between a token and a center. 
Figure~\ref{fig: bipartite} depicts an example of a bipartite graph to show the relationship of each token and centers. Solid lines show the edge with the highest weight for each token that represent the cluster that a token should be distributed to. Established techniques (\eg semantic graph \cite{LiuCroft}, Euclidean distance~\cite{aggarwal2019performance}) are to calculate the relatedness, however, these methods are not appropriate for encrypted tokens that are sparsely distributed~\cite{aggarwal2019performance}~\cite{LiuCroft}. 

As encrypted tokens lose their semantics, we ought to define the relatedness between tokens based on their statistical characteristics and then leverage it to distribute each token to the cluster that offers the maximum relatedness. 

Intuitively, the relatedness measure between tokens $t_i$ and $t_j$, denoted $r(t_i,t_j)$, is defined based on the magnitude of their \emph{co-occurrences}, \ie the number of documents where the two tokens appear together \cite{S3BD,clustcrypt}. Let $F_i$ and $F_j$ respectively denote the sets of documents that $t_i$ and $t_j$ are appeared in. Then, the intuitive co-occurrence of the two tokens is $F_{co}=F_i \cap F_j$. However, a deeper analysis reveals that quantifying the relatedness only based on the cardinality of co-occurrence (\ie $|F_{co}|$) can be misleading for the two following reasons:

\emph{First}, intuitive co-occurrence ignores the magnitude of disparity across $F_i$ and $F_j$ that negatively impacts the relatedness between $t_i$ and $t_j$. The disparity is determined based on the symmetric difference (\ie we have $F_{dis} = F_i \oplus F_j$). Accordingly, to consider the impact of both co-occurrence and disparity, we define a new measure, called \emph{relative co-occurrence}, and leverage it to determine the relatedness between $t_i$ and $t_j$. 

\emph{Second}, intuitive co-occurrence ignores the importance of $t_i$ and $t_j$ in each document $d\in F_{co}$. Accordingly, to measure the co-occurrence value in each document $d$, denoted $\upsilon(t_i,t_j,d)$, we consider the importance of each one of the tokens relative to their importance across all documents of $F_{co}$. We use frequency of a token in a document to measure its importance in that document. Formally, in document $d$, we calculate the value of co-occurrence based on Equation~\ref{eq:posCor}.

\begin{equation} \label{eq:posCor}
\vspace{-5pt}
\upsilon(t_i,t_j,d) = \frac{f(t_i,d)}
{ \sum\limits_{\forall m \in F_{co}}f(t_i,m)} \cdotp  \frac{f(t_j,d)}
{ \sum\limits_{\forall m \in F_{co}}f(t_j,m)}
\end{equation}

Similarly, we utilize Equation~\ref{eq:negCor} to measure the impact of disparity between two tokens in each document $d\in F_{dis}$, denoted $\varphi(t_i,t_j,d)$.

\begin{equation} \label{eq:negCor}
\vspace{-5pt}
\varphi(t_i,t_j,d) = \frac{f(t_i,d)}
{ \sum\limits_{\forall m \in F_{dis}}f(t_i,m)} +  \frac{f(t_j,d)}
{ \sum\limits_{\forall m \in F_{dis}}f(t_j,m)}
\end{equation}

In document $d$, once we know the co-occurrence and disparity between $t_i$ and $t_j$, we can calculate the relative co-occurrence as $\rho(t_i,t_j,d)=\upsilon(t_i,t_j,d)-varphi(t_i,t_j,d)$. Then, the relative co-occurrence across all documents of the two tokens (\ie $F_i \cup F_j$) is leveraged to calculate the relatedness between them.   

Assuming $c$ as the token that represents center of a given cluster (\ie $t_i=c\in centers$), we define relatedness between $c$ and token $t$, according to Equation~\ref{eq:distance}. Token $t$ is distributed to the cluster whose center offers the maximum relatedness. Note that, in this equation, to emphasize the importance of token $t$ in document $d$, we also consider its frequency ratio.

\begin{equation} \label{eq:distance}
\vspace{-5pt}
r(c, t) = \sum_{d\in (F_t \cup F_c)} \rho(t,c,d)\cdotp \frac{f(t,d)}{\sum\limits_{\forall m \in F_t}{f(t,m)}}  
\end{equation}

\subsection{Pruning Clusters to Expedite the Search Operation}
\label{cluster-abs}
The purpose of building topic-based clusters is to achieve scalable search over big data via limiting (pruning) the search scope based on the query topic, instead of exhaustively traversing the whole index structure. For pruning, we need to identify the clusters that are semantically relevant to the search query and discard the irrelevant ones. However, pruning is a challenging task when we operate on the encrypted data in the cloud. 

To overcome the challenge, we require the topic of each cluster in plain-text, such that we can identify the clusters whose topics are semantically related to the search query and only consider those clusters for searching. For that purpose, in our previous work \cite{sahan}, we established a method to represent the topic of each cluster $C_x$ (denoted $\alpha_x$) by considering the top-n most-frequent tokens of $C_x$. The tokens of $\alpha_x$ are decrypted and maintained on the edge tier of \name~in a structure called \emph{Abstract}.
Abstracts are leveraged to measure the topic similarity between a query and their corresponding clusters. In the next step, the search is conducted on the clusters that are most relevant to the query. For further details about creating abstracts and pruning operation, interested readers can refer to our earlier study~\cite{sahan,S3BD}.

\section{Privacy-Preserving Clustering Scheme for Dynamic Big Datasets (Dynamic \name)}
\label{sec:dynamic}
\subsection{Overview} 
In the previous section, we explained clustering of static (\eg archive) encrypted big datasets. However, many big datasets are dynamic (\eg healthcare data, criminal records) ~\cite{mary2012density} and their contents change over time. In this section, we deal with clustering and subsequently searching over such datasets. We consider two types of dynamic datasets: First is the \emph{semi-dynamic} datasets whose contents are updated in batch over time (\eg Museum of Modern Art (MoMA) dataset \cite{moma21}); Second is \emph{fully-dynamic} datasets whose contents are constantly updated (\eg Twitter streams \cite{twitter}).

The latest changes on the dataset have to be reflected in the clusters. Otherwise, altered documents are not retrieved by the search system, even if they include relevant contents. In fact, the updates on the dataset affect the tokens' co-occurrences and, subsequently, the clustering arrangement. As such, the challenge is to know \emph{how the addition or deleting documents change the topics and number of clusters}. 

Given the size of big datasets, reconstructing clusters (called \emph{re-clustering}) upon arrival of every single document or a small batch of documents is time-prohibitive. Moreover, the small updates generally cause negligible changes in the co-occurrences of tokens that are unlikely to modify the arrangement of clusters. Only significant updates can cause decisive changes on the magnitude of co-occurrence and relatedness that entail re-clustering. Accordingly, the two followup questions are: \emph{when to perform re-clustering?} and \emph{how to re-cluster the tokens?} 
To address these questions, based on the type of dynamic datasets, we propose two clustering schemes in \name: Semi-dynamic data clustering scheme (SD-\name) and Fully-dynamic data clustering scheme (FD-\name).

\subsection{Semi-Dynamic Data Clustering Scheme (SD-\name)}
\label{subsec:sdcs}
In semi-dynamic datasets, topic-based clustering can be initially achieved on the first batch of documents in the dataset according to the method described in the previous section. Then, the re-clustering decisions are made depending on the changes caused by the new batch of documents. That is, we need to determine whether the change caused by the  extracted tokens of the new batch is significant or not.

To determine the significance of changes caused by the tokens of the new batch, we utilize ${\chi}^2$ (chi-square) distribution test \cite{chil19} that can identify significant changes observed in a variable of a given population. The ${\chi}^2$ test is known as \emph{testing goodness of fit} and it is represented by Equation~\ref{chi}, where $O_i$ is the observed and $E_i$ is the expected value of a particular variable in $K$ trials.

\begin{equation}
\label{chi}
{\chi}^2 = \sum \limits_{ i=1} ^{ k} [ (O_i - E_i)^2 / E_i ]
\end{equation}

We consider the number of the extracted tokens in the new batch and the number of tokens in the existing clusters. Our null hypothesis ($H_0$) is to perform re-clustering and ${\chi}^2$ test is employed to check the validity of $H_0$. If the difference between the number of new tokens and existing tokens is small, a low value of ${\chi}^2$ is obtained. For one degree of freedom with $95\%$ confidence interval, the value of ${\chi}^2=3.841$ fails to reject $H_0$.
Alternatively, if the number of tokens in the new batch is significantly smaller than the number of existing tokens, ${\chi}^2$ value becomes higher that denotes significant deviation from ${H_0}$. Then, the decision is to reject ${H_0}$ and keep the existing clusters. 

Once the re-clustering decision is made, we use the method explained in Section~\ref{sec:static} to cluster tokens of the updated dataset. In the event that re-clustering is not achieved, the new tokens are accumulated with the of tokens of the next batches. As a result, the total number of new tokens becomes significant that leads to a lower ${\chi}^2$ value and subsequently acceptance of $H_0$.


\paragraph*{Updating Clusters}
Let $U_1$ a new batch of documents that introduces a set of new tokens $T=\{t_1, t_2,..., t_n\}$ that does not exist in the existing clusters. Assume that based on the re-clustering decision method, mentioned in the previous part, we determine to keep the existing clusters $\{C_1, C_2,..., C_n\}$ to accommodate $T$. 

To distribute $t_i\in T$ to a cluster, we can measure the relatedness as explained in Section~\ref{cluster-wise}. Alternatively, we can leverage the set of abstracts $\{A_1, A_2,..., A_n\}$. As they are in the plain-text format, a more accurate relatedness measurement can be conducted using the semantic similarity, as opposed to inferring the relatedness based on token co-occurrences in documents. In this case, we use Word2Vec~\cite{mikolov2013efficient} model to calculate the relatedness of $t_i$ and abstract $A_j$. Then, $t_i$ is assigned to a cluster that offers the highest relatedness. To avoid poor assignments, we define $\theta$ as the relatedness threshold that should be reached to assign $t_i$ to $C_j$. In the event that $t_i$ cannot join any cluster, a new cluster, called $C_{new} \in C$, is formed and $t_i$ is considered as its center. The above procedure is repeated for all $t_i \in T$.
 
\begin{algorithm}
	\SetAlgoLined\DontPrintSemicolon
	\SetKwInOut{Input}{Input}
	\SetKwInOut{Output}{Output}
	\SetKwFunction{algo}{algo}
	\SetKwFunction{proc}{Procedure}{}{}
	\SetKwFunction{main}{\textbf{ChooseCenter}}
	\SetKwFunction{quant}{\textbf{CalculateUniqueness}}
	\Input{set of abstracts $A$,  $tempIndex$ 
	, $\theta$}
	
	\Output{$H$, map of new tokens to clusters}
	
	\SetKwBlock{Function}{Function \texttt{ SD-\name($A, tempIndex, \theta$):}}{end}
	\Function{
			
			$T \gets tempIndex \setminus Central Index$ \;
			$H \gets \emptyset$ \;
			$A \gets \cup_{i=1}^{n}A_i$ \;
			$\Phi\gets \emptyset$\; //\small{Max-heap to find the abstract with highest similarity} \;
			\ForEach{token $t \in T$} {
				\ForEach{$a_{ij}\in A$}{
				$s \gets$ sim $(a_{ij}, t)$ \;
				    \If { $s > \theta$} {
    					Add $(s,i)$ to $\Phi$ \;
					}
				}
				\If {$\Phi \neq \emptyset$} {
				 //\small{Allocate $t$ to existing cluster\;
				 $(t,i) \gets$ Extract max pair from $\Phi$ \;
					Add $(t,i)$ to $H$\;}
				
				}
				
				\Else {
				//\small{Forming a new abstract and cluster and add it to $H$}\;
				    $A_{n+1} \gets \{t\}$ \;
			        $A \gets \cup_{i=1}^{n+1}A_i$ \;
					Add $(t,n+1)$ to $H$\;
				}
			}
		Encrypt $H$ and push it to the cloud tier \;
		
	}

	\caption{Pseudo-code to update clusters in SD-\name.}
	\label{alg:semi}  
      
\end{algorithm}

\paragraph*{Determining the value of $\theta$ Threshold}
We estimate the value of $\theta$ threshold by leveraging the abstracts $\{A_1, A_2,...A_n\}$. Recall that the elements of abstract $A_i$ are the ones that best represent the topic of its corresponding cluster $C_i$. We define coherency of $A_i$ as the average similarity distance across pairs of its elements. Let $\{a_{i1},..., a_{ip}\}$ be the set of elements of $A_i$. Then, coherency of $A_i$, denoted $K_i$, is defined based on Equation~\ref{eq:koh} where $sim(x,y)$ shows the similarity distance between $(x,y)\in A_i\times A_i$.

\begin{equation}\label{eq:koh}
    K_i=\frac{\sum\limits_{\forall (x,y)\in A_i\times A_i | x\neq y} Sim(x,y)}{\binom p2}
\end{equation}

Then, we define $\theta$ as the global minimum across all abstracts (\ie $\theta=\min_{\forall i} K_i$). This implies that a new token can join a cluster only if its distance does not worsen the coherency of current clusters. 
Otherwise, the new token forms its own cluster.  

Algorithm~\ref{alg:semi} shows the pseudo-code of how to update clusters in SD-\name, in case we choose not to perform re-clustering. In addition to the set of abstracts ($A$) and $\theta$, the algorithm receives the set of tokens for a new document batch, which is stored in form of a temporary index. 
The algorithm returns the $H$ structure that includes the mapping of new tokens to their respective clusters. In Steps $7-9$, for each new token, we calculate the similarity distance with respect to all abstract elements $a_{ij}$ and check whether the similarity distance exceeds $\theta$ or not. If it exceeds $\theta$,
we make a pair of similarity distance and corresponding abstract number, denoted as $(a_{ij},t)$ and build max-heap $\Phi$ based on the distance (in Step $10-12$). If $\Phi$ contains any value, we extract from it the pair that has the largest value (\ie the abstract that offers the most topic similarity for $t$). Then, in Step 17, the pair of $(t,i)$ is added to $H$. 
On the contrary, if $\Phi$ is null, it implies that no cluster offers a considerable similarity to $t$, and so, in Steps $19-24$, we build a new abstract and cluster using $t$.  
Finally, we encrypt the tokens of $H$ and push it to the cloud tier. On the cloud end, \emph{cluster manager} updates its clusters based on $H$.    


\subsection{Fully-Dynamic Data Clustering Scheme (FD-\name)}
\label{subsec:fdcs}
Unlike SD-\name, for fully-dynamic datasets, clusters have to be formed or updated upon arrival of the documents. That is, continuous or burst arrival of new documents should trigger FD-\name. Accordingly, in FD-\name, we consider two cases in forming clusters: (A) \emph{initial case} that occurs when first document arrives and there is no existing cluster and (B) \emph{update case}, where the existing clusters have to be updated based on the new changes in the dataset. 

In the initial case, the edge tier extracts the set of new tokens from the uploaded document(s).
We designate the token with the highest frequency to represent the topic and choose it as the cluster center too. 
Then, the second most frequent token is clustered based on its similarity distance with the designated cluster center, according to the method discussed in Section~\ref{subsec:sdcs}. Also, to determine joining the existing cluster or forming a new one, we initialize the threshold to $\theta = 0.1$. This procedure continues until all tokens are clustered. In the update case, we apply the same method as SD-\name. That is, upon uploading a document, the system decides to either perform re-clustering or updating existing clusters. 

\section{Security Analysis}\label{sec: seca}
The proposed clustering schemes are applicable in the context of searchable encryption and document retrieval systems. According to the three-tier architecture, described in Figure \ref {sd}, client- and edge tiers are in the user premises, hence, the activities conducted and the user's key on these tiers are considered safe and trusted. The \emph{Abstract} structures are kept on the edge tier in plain-text to enable us to measure the similarity with the search phrase and performing pruning. 

On the other hand, activities performed on the cloud-tier are considered as dishonest and prone to different types of attacks. We are concerned about both internal (\ie affiliated parties) and external (\ie unaffiliated outside intruders) attackers who desire to learn the encrypted clustered tokens and documents. To explain the threats of the attackers, we provide the following preliminaries:

\textit{View}: This term denotes the portion that is visible to the cloud during any given interaction among client, edge, and server. The central index and the set of clusters {$C_1...C_n$}, the trapdoor of the given encrypted search query $Q^'$, and the collection of encrypted documents $D^'$. In some models, $Q^'$ also contains a particular weight for each term. The search results related to $Q^'$ are considered as $I_c$. The view of expanded $Q^'$ and $I_c$ are symbolized as $V(Q^')$ and $V(I_c)$ respectively.

\textit{Trace}: This term denotes the information exposed about $I_c$. Our aim is to allow the attacker to infer the information of $I_c$ as little as possible.  

The View and Trace enclose all the information that the attacker would gain. To encrypt the document set we use probabilistic encryption model that is considered to be one of the most secure encryption techniques~\cite{S3BD,cryptdb}. This does not utilize one-to-one mapping and so, $D^'$ is not prone to dictionary-based attacks~\cite{dictionaryAttacks}. Each token in a cluster is deterministically encrypted. Thus, each cluster in the View, only shows an encrypted mapping of the tokens and their co-occurrences in the plain-text format.

If any type of attacker can gain access to the cloud, he/she could only understand the importance of a particular encrypted token by observing the co-occurrences. It is technically possible to encrypt co-occurrences using homomorphic encryption \cite{homomorphic:modern} and perform computation on the co-occurrences while it is in the encrypted form. However, in Section~\ref{sec-background}, we discuss that this technique practically falls short on performance~\cite{Naehrig:2011} and affects the real-time behavior of the search system. As such, in the current implementation, we use co-occurrence information in the plain-text format. Note that, even when the co-occurrences are not encrypted, the attacker cannot decrypt the token. 

An attacker could obtain a Trace regarding $V(Q^')$. From that view, the attacker could only understand the importance of each search term from $Q^'$ by analyzing the associated weights of the query terms. Similar to the previous consideration, the attacker is not able to reveal the search terms from $Q^'$. In spite of a minimally trusted computing base, an attacker may still intend to access the system through
man-in-the-middle, either \emph{honest but compromised} or \emph{untrusted} cloud providers to attack the confidentiality of the user data.
By any means, if the attacker successfully performs a man-in-the-middle attack, he/she can access the document list $V(I_c)$ resulting from searching $Q^'$ with Trace. At this point, the attacker may only obtain the documents' names with encrypted contents that are unreadable.

There are methods (e.g., \cite{george2021structured}) that can be used to tackle
frequency attacks when the searches and cluster updates are predictable.
Theoretically, an attacker could build a dictionary considering all the clusters' tokens by performing frequency attack. Eventually, the attacker tries to build a clone document set $D'$ utilizing the dictionary. Although all of the tokens extracted from a particular document are sufficient to learn the topic of the document, it is not possible to unveil the whole document as we do not use all of the keywords of the document set to build the encrypted index. Besides, we encrypt the whole document at once instead of word level encryption before outsourcing it to the cloud. This procedure ensures that even if the document set is compromised on the cloud tier, it is impossible to perform a dictionary attack.

Even if the attacker knows the trace, he/she cannot understand what exactly the retrieved encrypted documents convey. Moreover, attacks can be occurred in the communication between the edge and cloud tiers. In this case, by monitoring the search process, an attacker could obtain the resultant document list for $Q'$. However, the attacker is not able to decrypt the documents, since they can be decrypted only when they are downloaded on the edge system.  

An attacker could also attempt to modify data (\eg encrypted tokens and documents) in the clusters.  Such attacks can potentially tamper with the integrity of user data. However, this type of attack could be detected, because neither the edge will be able to decrypt the modified tokens to form or update \emph{Abstracts}, nor the user will be able to decrypt the retrieved documents in the original plain-text form. This is because of applying symmetric encryption (\eg AES encryption) on the user’s data with keys managed by the user. Hence, in the event that
the encrypted data are altered by an attacker, such data cannot be decrypted by the users' keys.  
Actually, protecting the user's key is crucial to restrain possible attacks. If the key is compromised, the system cannot detect the attacker and, therefore, both tokens and documents can be exposed.

\section{Performance Evaluation}\label{sec:evltn}
\subsection{Experimental Setup}
We developed a working version of ~\name~and made it available publicly in our Github\footnote{\url{https://git.io/fjDsq}}.
We evaluate the performance of \name~using three distinct datasets that have different properties and volumes. We compare and analyze the clustering quality with other approaches that operate in encrypted or unencrypted domains. The experiments were conducted on a machine with two $10$-core $2.8$ GHz E5 Intel Xeon processors
and $64$ GB of memory. 

To evaluate the performance of \name~in handling big data, we used a subset of \texttt{Amazon Common Crawl Corpus (ACCC)} dataset (available in~\cite{comcrawl}). The whole dataset size $\approx$ $150$ terabytes that contains different web-based contents, such as blogs and social media contents. We randomly selected $6,119$
documents that collectively form a $\approx$ $500$ GB document set. The second
dataset, named \texttt{Request For Comments (RFC)} (available in~\cite{rfc}), is domain-
specific and includes documents about the internet and communication networks. RFC includes $2,000$ documents and its total size is
$\approx$ 247 MB. The third dataset is \texttt{BBC} (available in~\cite{bbc}) that is not domain-specific and includes news in certain categories such as technology, politics, sports, entertainments, and business. It contains $2,225$ documents and is $\approx$ $5$ MB. The reason for choosing this small
dataset is that, unlike ACCC and RFC, each document of BBC is short and we can verify clusters’ coherency manually. For each dataset, the documents are passed through Maui keyword extractor~\cite{Maui} to identify keywords semantically represent the document.


\subsection{Evaluation Metrics and Baselines from Prior Works}

For performance evaluation of \name, we compare it against five other schemes, where two schemes cluster plain-text data and the other three schemes cluster encrypted data. Among the two, one of the schemes \emph{W2V Kmeans}) is based on \textit{K}-means clustering~\cite{kmeans} where feature extraction is done based on Word2Vec~\cite{mikolov2013efficient} embedding.

Another scheme, \textit{WordNet}~\cite{millerwordnet}, is an enhanced version of \textit{K}-means that generates synonym set based on the input data and then, applies $K$-means clustering on the sets.
Token distribution in WordNet is performed based on edge counting method, proposed by Wu and Palmer~\cite{millerwordnet}. 


Three encrypted clustering schemes that have been used in the comparison are namely, S3BD~\cite{S3BD},~\textit{HK}-means++~\cite{wang2019research}, and ClustCrypt~\cite{clustcrypt}). We have discussed S3BD and~\textit{HK}-means++ in Section~\ref{sec-background}.
 \emph{ClustCrypt} is the preliminary version of S-\name. Their difference mainly lies in the way tokens are distributed across the clusters. In ClustCrypt, the relatedness is simply calculated based on contribution and co-occurrences metrics, whereas in S-\name, the magnitude of both similarity and disparity are considered to measure the relatedness (see Section~\ref{cluster-wise} for further details).

The goodness of clusters set can be quantified by a number of evaluation metrics. However,
evaluating the performance of a clustering scheme is not as simple as counting errors in classification algorithm. Specifically, instead of considering the absolute values of cluster labels, cluster evaluation metrics either measure the separation of clustered data similar to ground truth set of classes or internal cluster validation. Internal cluster validation denotes that members belong to the same class should be more similar than members of other classes and vice versa.  
In practice, class label information is not always available in most of the application scenarios and, therefore, internal validation metrics are the only option for validation in such situation~\cite{rodriguez2019clustering,kwon2017clustervision}.

As there is no ground truth for the considered datasets, we choose evaluation metrics that evaluate the clusters based on statistical analysis of the cluster members. We evaluate three widely-adopted clustering metrics, namely Silhouette coefficient (SC), Calinski-Harabasz index (CI), and Davies-Bouldin index (DI). 

\textit{Silhouette Coefficient (SC)} score interprets and validates intra-cluster consistency. In particular, the metric signifies how similar a cluster member is to its own cluster compared to the other clusters. The value of the SC score ranges from $-1$ to $+1$, where a high value indicates that a given member is well matched to its own cluster and poorly matched to the other ones. \textit{Calinski-Harabasz Index (CI)} denotes how well-defined (\ie well-separated) the clusters are. The CI value of clusters is calculated based on the ratio of the sum of between-clusters dispersion to the sum of inter-cluster dispersion. A higher CI value indicates a more topically separated (\ie less overlapping) clustering and vice versa. Similar to the CI metric, \textit{Davies Bouldin Index (DI)} is used to measure the goodness of separation across clusters and the reason we consider it in our evaluation is to verify the CI metric evaluation for the clusters. DI is calculated based on the ratio of within-cluster distances to the between-cluster distances. A lower DI value indicates a more topically-separated clustering and it is preferred. In addition to these metrics, we measure the \emph{clusters' coherency} to evaluate the quality of the topic-based clustering within each cluster. This is a similarity-based evaluation metric to calculate the average of all possible pair-wise token similarity for a given cluster. In fact, Coherency represents how the tokens in a cluster are related to a certain topic. Then, the average of coherency across all clusters is calculated to represent the overall quality of a certain clustering method.

We instrument the pre-trained Google News Word2vec model~\cite{mikolov2013efficient} to determine the similarity between any two given keywords. The model is a $300$-dimension vector representation of three million phrases.
The model requires a text dataset as input to build a vocabulary from the input dataset and learns vector representation of the words in the dataset. The model uses cosine similarity and provides the score ($-1\leq similarity\; score\leq 1$) for any two given tokens. We note that, the pre-trained Word2vec model operates only on plain-text tokens. Subsequently, we do not encrypt the tokens while uploading for evaluation purposes. However, the proposed schemes assume tokens to be encrypted and do not use the properties of plain-text tokens.

\subsection{Evaluation Results}
\label{subsec:evaluationres}
\subsubsection{Evaluating Silhouette Coefficient (SC) Score\nopunct}\hfill\\

\begin{figure} 
 	\centering
 	\includegraphics[width=\linewidth]{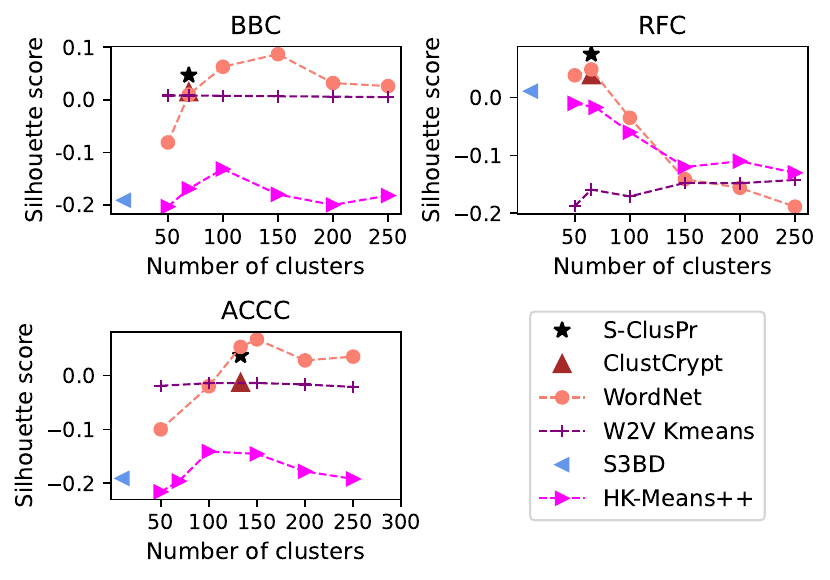}
 	\caption{\small{Silhouette Coefficient (SC) metric for each dataset. The results are obtained from S-\name, \textit{HK-}means++, ClustCrypt (that are encrypted-based clustering schemes), W2V-Kmeans, and WordNet clustering schemes (that operate on plain-text tokens). }}
 	\label{fig: silh}
 \end{figure}

Figure~\ref{fig: silh}
shows the results of SC score evaluation on the three datasets and for varying number of clusters (in the horizontal axis). We note that, for this experiment, the value of $K$ in W2V Kmeans, WordNet, and \textit{HK-}means++ is randomly chosen and iteratively evolves. As such, we calculate the SC score for all the considered $K$ values and show them in multiple data points in the figure. However, other schemes (namely, S-\name, ClustCrypt, S3BD) are not iterative and provide only one SC score for their determined $K$ values.

As the procedure of estimating the number of clusters is similar in ClustCrypt and S-\name~schemes, we can see that both of the schemes generate $69$, $65$, and $133$ clusters for the \texttt{BBC}, \texttt{RFC}, and \texttt{ACCC} datasets, respectively. As \texttt{ACCC} is the largest 
and broadest (\ie not domain-specific) dataset, it yields the highest $K$ value. \texttt{RFC} is not the smallest dataset, however, due to its domain-specific nature, it yields the lowest $K$ value.

Figure~\ref{fig: silh} represents SC metric outcomes for S-\name and the five other compared schemes. According to the figure, considering all of the datasets, overall top performers are 
WordNet and S-\name. Moreover, S-\name~outperforms others in the \texttt{RFC} dataset. On the contrary, \textit{blue}{ \textit{HK-}means++ and S3BD underperform in most of the situation.}       
The experiment indicates that the cluster sets generated by \textit{HK}-means++ and S3BD contain less intra-cluster similarity. 
WordNet and S-\name~provide the highest intra-cluster similarity and hence, outperform others in all datasets.

  
 \begin{figure} 
 	\centering
 	\includegraphics[width=\linewidth]{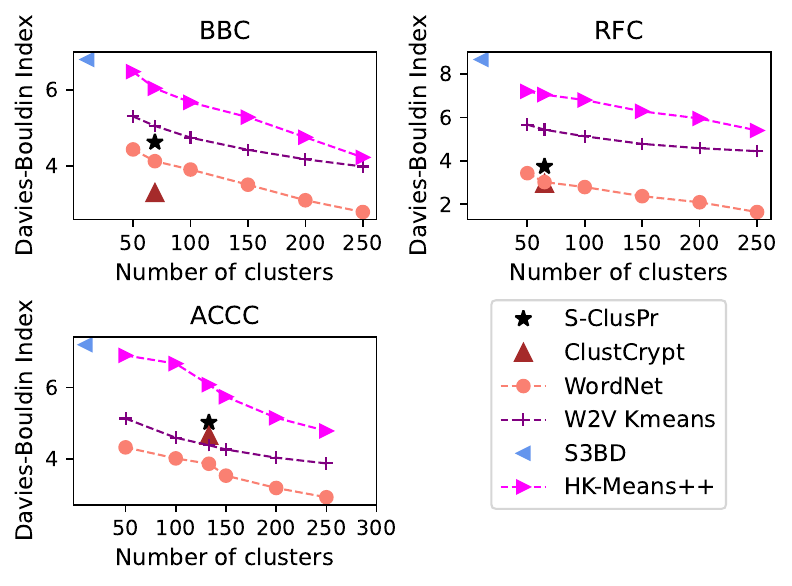}
 	\caption{\small{Davies-Bouldin Index (DI) for each dataset using different clustering schemes.}}
 	\label{fig: boud}
 \end{figure}

 \begin{figure} 
 	\centering
 	\includegraphics[width=\linewidth]{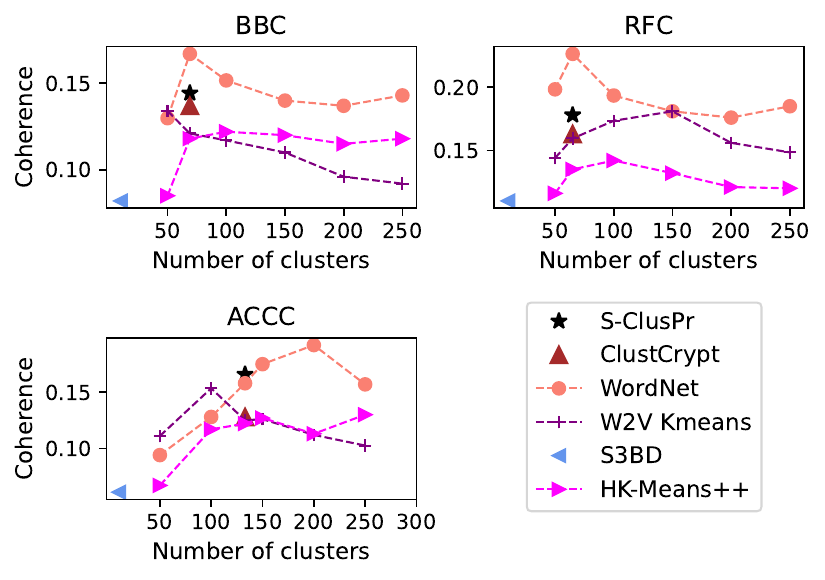}
 	\caption{\small{Cluster coherency for each dataset.}}
 	\label{fig: main}
 \end{figure}

\begin{table}[h]
\begin{subtable} {\linewidth}
\label{cal:bbc}
\subcaption{BBC}
\centering
\resizebox{\linewidth}{!}{
\begin{tabular}{c|c|c|c|l|c|c|}
\cline{2-7}
                                                                                & \multicolumn{6}{c|}{Approaches}                                                                                                                                                                                                                                                  \\ \hline
\multicolumn{1}{|c||}{\begin{tabular}[c]{@{}c@{}}No. of \\ Cluster\end{tabular}} & \begin{tabular}[c]{@{}c@{}}HK- \\means++\end{tabular} & WordNet        & \begin{tabular}[c]{@{}c@{}}W2V \\ Kmeans\end{tabular} & S3BD                                  & ClustCrypt  & S-ClusPr             \\ \hline \hline
\multicolumn{1}{|c||}{10}                                                        &-                                                          &-                                                                               & -                                                      & \ \ \ \ 8.7                                          & \ -   & \ -                    \\ \hline
\multicolumn{1}{|c||}{50}                                                        & 25.43                                                   & \textbf{277.53}                                                                     & 11.16                                                 & \ \ \ \ -                                               & \ -   & \ -                   \\ \hline

\multicolumn{1}{|c||}{69}                                                        & 18.47                                                   & \textbf{253.60}                                                                     & 9.22                                                  & \ \ \ \ -                                             & 11.70  & 13.58                 \\ \hline
\multicolumn{1}{|c||}{100}                                                       & 11.13                                                    & \textbf{203.87}                                                                   & 7.37                                                  & \ \ \ \ -                                               & -  & \ -                    \\ \hline
\multicolumn{1}{|c||}{150}                                                       & 14.05                                                    & \textbf{164.43}                                                                    & 5.81                                                  & \ \ \ \ -                                               &-    & \ -                   \\ \hline
\multicolumn{1}{|c||}{200}                                                       & 10.17                                                    & 122.51                                                                     & 4.93                                                  & \ \ \ \ -                                               & -       & \ -               \\ \hline
\multicolumn{1}{|c||}{250}                                                       & 12.02                                                    & 97.15                                                                      & 4.38                                                  &  \ \ \ \ -                                              &  -      & \ -               \\ \hline
\end{tabular}
}
\end{subtable}

\bigskip

\begin{subtable} {\linewidth} 
\subcaption{RFC}
\centering
\resizebox{\linewidth}{!}{
\begin{tabular}{c|c|c|c|l|c|c|}
\cline{2-7}
                                                                                & \multicolumn{6}{c|}{Approaches}                                                                                                                                                                                                                                                             \\ \hline
\multicolumn{1}{|c||}{\begin{tabular}[c]{@{}c@{}}No. of \\ Cluster\end{tabular}} & \begin{tabular}[c]{@{}c@{}}HK- \\ means++\end{tabular} & WordNet                       & \begin{tabular}[c]{@{}c@{}}W2V \\ Kmeans\end{tabular} & S3BD                                 & ClustCrypt      & S-ClusPr              \\ \hline \hline
\multicolumn{1}{|c||}{10}                                                        & -                                                         & -                                                                                       & -                                                      & 1247.20                                            & -    & -                  \\ \hline

\multicolumn{1}{|c||}{50}                                                        & 1730.26                                                  & 4320.63                                                                          & \textbf{60380.05}                                              & \ \ \ \ -                         & \ -                & \ -               \\ \hline

\multicolumn{1}{|c||}{65}                                                        & {1945.42}                             & { 3980.75}                                     & \multicolumn{1}{l|}{\textbf{51564.61}}                                 & \multicolumn{1}{l|}{\ \ \ \ -}         & 23760.64 & \textbf{29439.30}              \\ \hline
\multicolumn{1}{|c||}{100}                                                       & 1834.64                                                  & 3660.78                                                                           & \textbf{24374.17}                                              & \ \ \ \ -   & -                                                 & -                      \\ \hline
\multicolumn{1}{|c||}{150}                                                       & 1684.47                                                  & 3110.25                                                                          & \textbf{18684.33}                                              & \ \ \ \ -                           & -                          &  -                     \\ \hline
\multicolumn{1}{|c||}{200}                                                       & 846.71                                                  & 2572.89                                                                         & 16746.74                                              &  \ \ \ \ -                                                  & -  & -                     \\ \hline

\multicolumn{1}{|c||}{250}                                                       & 436.43                                                   & 1834.58                                                                          & 15139.11                                              & \ \ \ \ -                                                    & -   & -                   \\ \hline
\end{tabular}
}
\label{tab:cal rfc}
\end{subtable}

\caption{Calinski-Harabasz Index for the datasets.}
\label{tab:cal}
\end{table}


\subsubsection{Evaluating Calinski-Harabasz Index (CI)\nopunct}\hfill\\
Table~\ref{tab:cal} represents CI metric outcomes for S-\name~and the five other schemes. According to the table, the RFC clusters provide large CI values compared to the BBC dataset, regardless of the employed clustering scheme. It is noteworthy that, we had the same observation for the ACCC dataset, however, we do not show its table due to the shortage of space. The superiority of RFC is because it is a domain-specific dataset with a few topics compared to the other two. 
Within Table~\ref{tab:cal rfc}, we can see that although W2V-Kmeans significantly outperforms the other schemes for most of the $K$ values, WordNet, ClustCrypt, and S-\name~also provide satisfactory CI values that imply well-partitioned clusters.    
\subsubsection{Evaluating Davies Bouldin Index (DI)\nopunct}\hfill\\
The DI values for the clusters, obtained by S-\name~and the compared schemes are expressed in Figure~\ref{fig: boud}. In most of the scenarios, we observe that increasing the number of clusters reduces the DI value. This is because, typically,
configuring clustering schemes to build more clusters on a given dataset leads to a higher coherency within each of the clusters. 

According to the figure, we observe that WordNet scheme outperforms others. The DI value for S-\name~is in the acceptable range, which indicates that the scheme can offer a competitive goodness of separation across clusters in compared to the most of other schemes. On the other hand, higher DI value yielded by \textit{HK-}means++ signifies poor cluster separation.

\subsubsection{Evaluating Cluster Coherency\nopunct}\hfill\\
Figure~\ref{fig: main} shows the clusters' coherency on the three datasets using various clustering schemes. 
Using S-\name, $69$, $65$, and $133$ clusters are created for the \texttt{BBC}, \texttt{RFC}, and \texttt{ACCC} datasets, respectively. As \texttt{ACCC} is the largest 
and broadest (\ie not domain-specific) dataset, it yields the highest $K$ value. \texttt{RFC} is not the smallest dataset, however, due to its domain-specific nature, it yields the lowest $K$ value. For the same reason, across the three datasets, S-\name~offers the highest coherency value ($\approx 0.16$) for the \texttt{RFC} dataset.

In compare to ClustCrypt, we notice that S-\name~offers a negligible coherency improvement ($\approx 6\%$) for the \texttt{BBC} and \texttt{RFC} datasets. However, for the \texttt{ACCC} dataset, S-\name~improves the coherency by approximately $31\%$.    

Analysis of the plain-text-based schemes reveal that, WordNet clusters offer the highest coherency value. This is expected, because it is difficult for an encrypted clustering scheme (\eg S-\name) to outperform the unencrypted ones, since they do not have access to the semantics of the tokens~\cite{millerwordnet} to build the clusters. However, we observe that the coherency offered by S-\name~competes with the one offered by the $K$-means scheme. In particular, S-\name~provides a higher coherency value than $K$-means for the \texttt{RFC} and \texttt{BBC} datasets.

To evaluate the suitability of estimated number of clusters ($K$) by S-\name, we configure both $K$-means and WordNet to use the estimated $K$ number of clusters for the studied datasets. According to the figure, for \texttt{RFC} and \texttt{BBC}, S-\name~suggested sets of $K$ clusters offer a higher coherency than $K$-means and a comparable one to WordNet. In the case of \texttt{ACCC}, S-\name~even outperforms WordNet in terms of coherency.

\subsubsection{Analyzing the Impact of S-\name~on Searchable Encryption Systems}
\begin{figure}
	\centering
	\includegraphics[width=.7\linewidth]{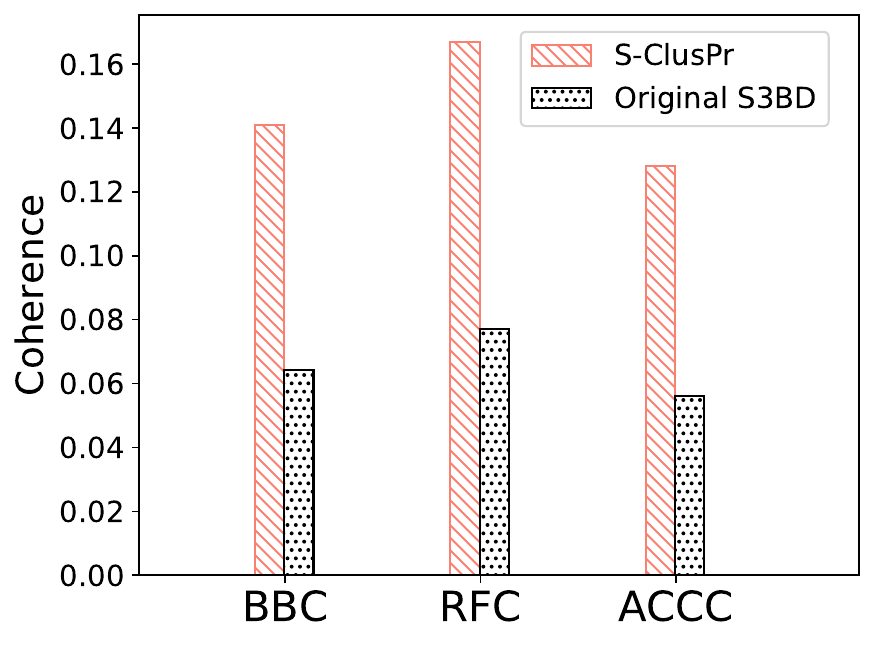}
	\caption{\small{Comparing the impact of clustering using S-\name~against original clustering of S3BD for the studied datasets.}}
	\label{fig: Pro_en}
\end{figure}
One objective of this research is to enhance the performance of S3BD secure search system. As such, we instrumented S-\name~in S3BD and compared the coherency of resulting clusters with its original clustering scheme that predetermines a value for $k=10$. Moreover, its center selection only considers the co-occurrences. In this experiment, we intend to evaluate the improvement that S-\name~achieves within S3BD on the three studied datasets. In this experiment, the estimated values of $K$ for \texttt{BBC}, \texttt{RFC}, and \texttt{ACCC} are $69$, $65$, and $133$, respectively.

\paragraph{a) Impact on the Clustering Coherency of S3BD.\nopunct}
Figure \ref{fig: Pro_en} shows that for all the studied datasets, clusters generated by S-\name~have remarkably higher coherency than the original clustering scheme of S3BD. This shows determining number of clusters based on dataset characteristics and choosing center tokens based on the centrality concept is effective. Our hypothesis is that, such efficiency improves the accuracy and offers more relevant semantic search results. This is because tokens of the clusters are more congruent to the clusters' topics, hence, more effective pruning is accomplished. For further evaluation of this hypothesis, next experiments concentrate on the impact of S-\name~on the search quality.

\begin{figure}
	\centering
	\includegraphics[width=\linewidth]{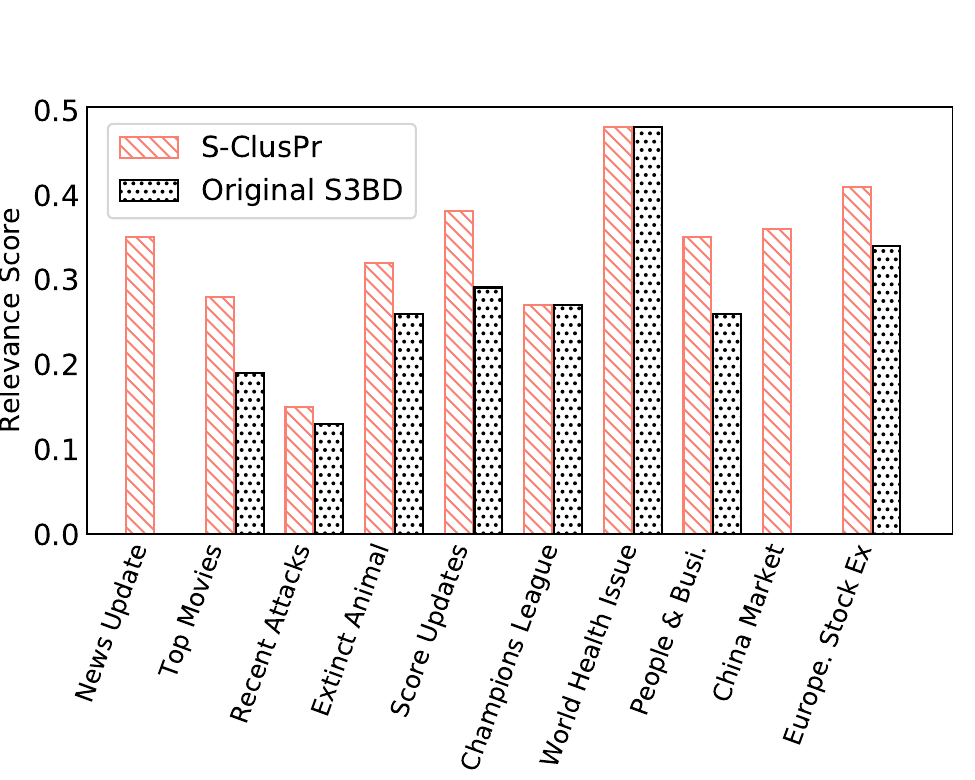}
	\vspace{-5pt}
	\caption{\small{Comparing the relevancy of search results using S-\name~vs original S3BD clustering in BBC dataset. The value of relevancy is calculated based on TSAP@10 scoring metric.}}
	\label{fig: search_ac}
	
\end{figure}

\paragraph{b) Impact on the Search Accuracy of S3BD}
The purpose of improving the clusters' coherency in this study is to ultimately enhance the search accuracy by retrieving more relevant documents. To evaluate the impact of such improvement, in this part, we compare and analyze how the search accuracy of S3BD system is affected by utilizing S-\name's clusters against the circumstance where its original clustering method is utilized.
For the evaluation, we generated a set of $10$ benchmark search queries that are listed in Table~\ref{bench}. 

\begin{table}
	
	\centering
	\resizebox{\linewidth}{!}{
	
	\begin{tabular} {|p{3.4cm}|p{3.1cm}|p{3.1cm}|}
		
		\hline
		
		{\texttt{\textbf{ACCC Dataset}}  }                  & {\texttt{\textbf{BBC Dataset}}}                & {\texttt{\textbf{RFC Dataset}}}                   \\ \hline
		
		\texttt{{Orlando Magic}}                  & \texttt{{News Update}}        & \texttt{{Internet}}              \\ \hline
		
		\texttt{{Samsung Galaxy}}         & \texttt{{Top Movies}}             & \texttt{{TCP}}                   \\ \hline
		
		\texttt{{Baseball routine}}        & \texttt{{Recent Attacks}}        & \texttt{{Fiber Doctor}} \\ \hline
		
		\texttt{{Recommendation}}                & \texttt{{Endangered Animals}} & \texttt{{Wifi}}                  \\ \hline
		
		\texttt{{North America}}     & \texttt{{Score Updates}}             & \texttt{{IoT}}    \\ \hline
		
		\texttt{{Tennis Tournament}} & \texttt{{Champions League}}       & \texttt{{Radio Frequency}}       \\ \hline
		
		\texttt{{Holy Martyr}}       & \texttt{{World Health Issue}}     & \texttt{{UDP}}                   \\ \hline
		
		\texttt{{Library}}            & \texttt{{People and Business }}         & \texttt{{Edge Computing}}        \\ \hline
		
		\texttt{{Stardock}}               & \texttt{{China Market}}          & \texttt{{Encryption Schemes}} \\ \hline
		
		\texttt{{Orthodox Church}}    & \texttt{{European Stock Exchange}} & \texttt{{Broadcasting}}          \\ \hline
		
	\end{tabular}
	}
	\caption{\small{Benchmark queries for each one of the studied datasets.}}
	
	\label{bench}
\end{table}


To measure the relevancy of search results for each query, we use \textit{TREC-Style Average Precision} scoring method~\cite{mariappan}. This method works based on the recall-precision concept and the score is calculated by $\sum_{i=0}^N r_i/N$, where $r_i$ denotes the score for $i^{th}$ retrieved document and $N$ is the cutoff number (number of elements in the search results) that we consider as $10$. Therefore, we call it \emph{TSAP@10}. 



We measure TSAP@10 score only for the \texttt{RFC} dataset and its benchmark queries. The reason is that it is domain-specific and feasible to determine the relevancy of the retrieved documents. 
To compare the relevancy provided by S-\name~against the original S3BD clustering, we apply the benchmark queries to the S3BD search system. In Figure~\ref{fig: search_ac}, the relevancy score of the results for each query when the two clustering schemes are applied are measured and presented.
According to the Figure, for most of the queries, S-\name~clustering offers a higher relevancy score. For the two queries that have identical \emph{TSAP@10} score, their retrieved document lists are equivalent. Also, S-\name~clusters provide score for \textit{News Update} and \textit{China Market} benchmark queries, whereas original S3BD clusters do not retrieve any relevant documents for these queries.

\paragraph{c) Impact on the Search Time of S3BD}

\begin{figure}
	\centering
	\includegraphics[width=.7\linewidth]{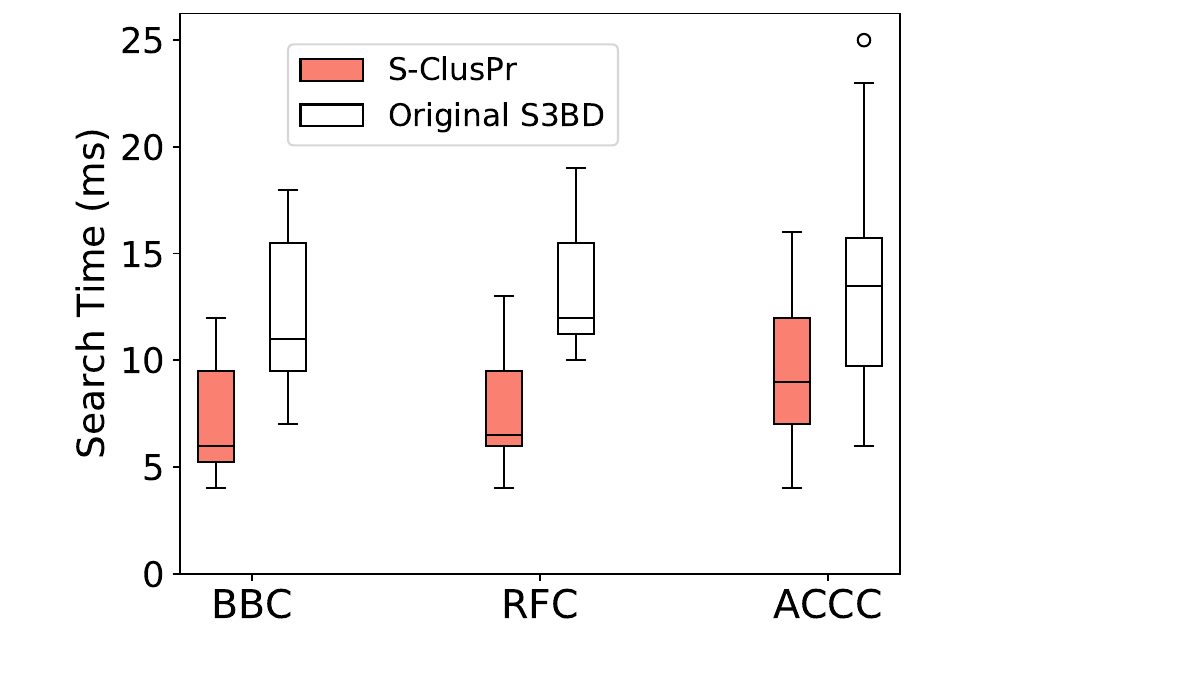}
	\caption{\small{Search time of S3BD when S-\name~is used for clustering versus when the original S3BD clustering is used.}}
	\label{fig: time_en}
\end{figure}

Figure~\ref{fig: time_en} presents the total search time of the benchmark queries for each dataset. The search time is measured as the turnaround time of searching each query---from the time a query is issued until the result set is received. To eliminate the impact of any randomness in the computing system, we searched each set of benchmarks 10 times and reported the results in form of box plots.
The figure indicates that  when S-\name~clustering is utilized, the search time is significantly shorter than the circumstance where the original S3BD clustering is used. Longer search time impacts the scalability and real-time quality of the search operation on big data. 
Analyzing Figures~\ref{fig: Pro_en} to~\ref{fig: time_en} reveals that integrating S-\name~in the search system, not only makes it more accurate, but makes it faster and more scalable too.   

\begin{figure*} [!htbp]
\begin{subfigure}{.33\textwidth}
\centering
\includegraphics[width=\linewidth]{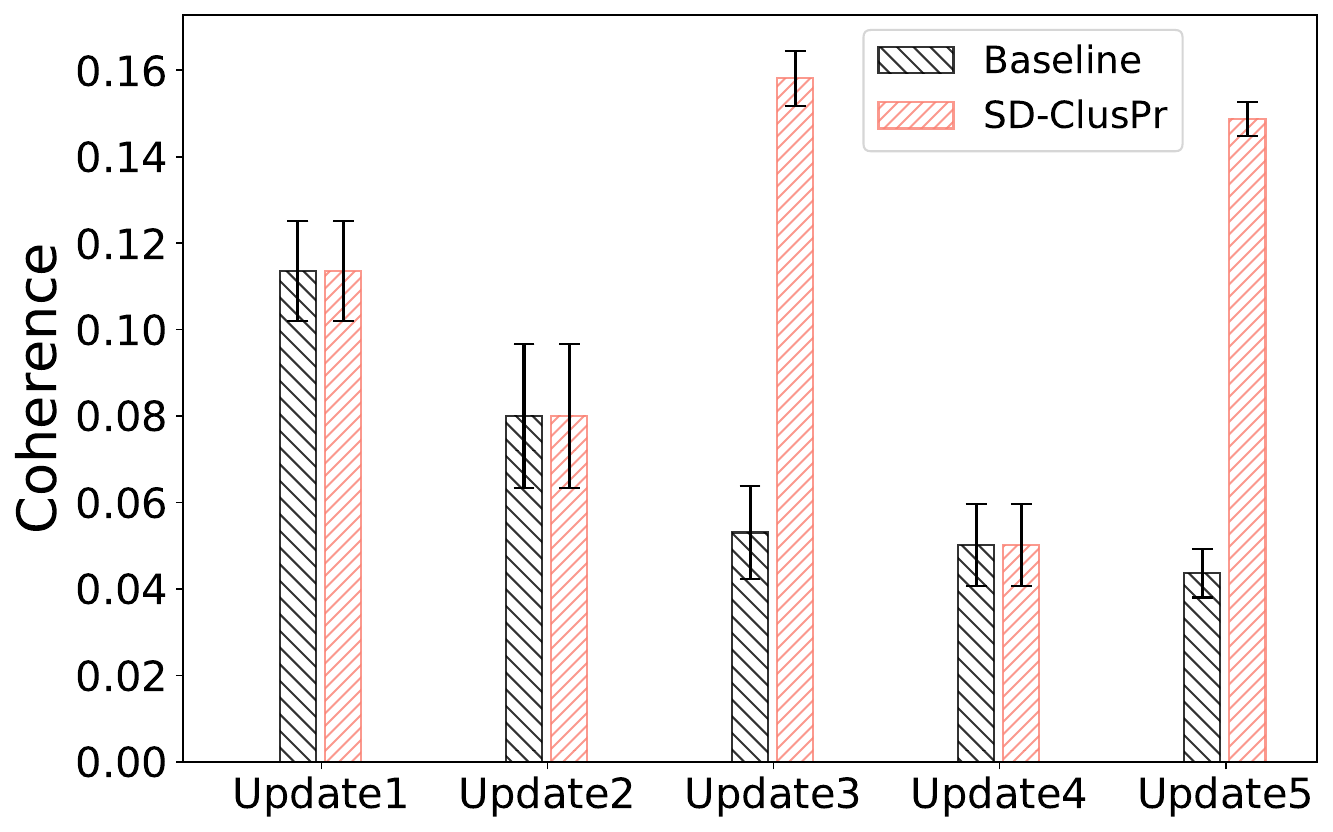}
\caption{BBC Dataset}
\label{fig: semidyna_bbc}
\end{subfigure}\hfill
\begin{subfigure}{.33\textwidth}
\centering
\includegraphics[width=\linewidth]{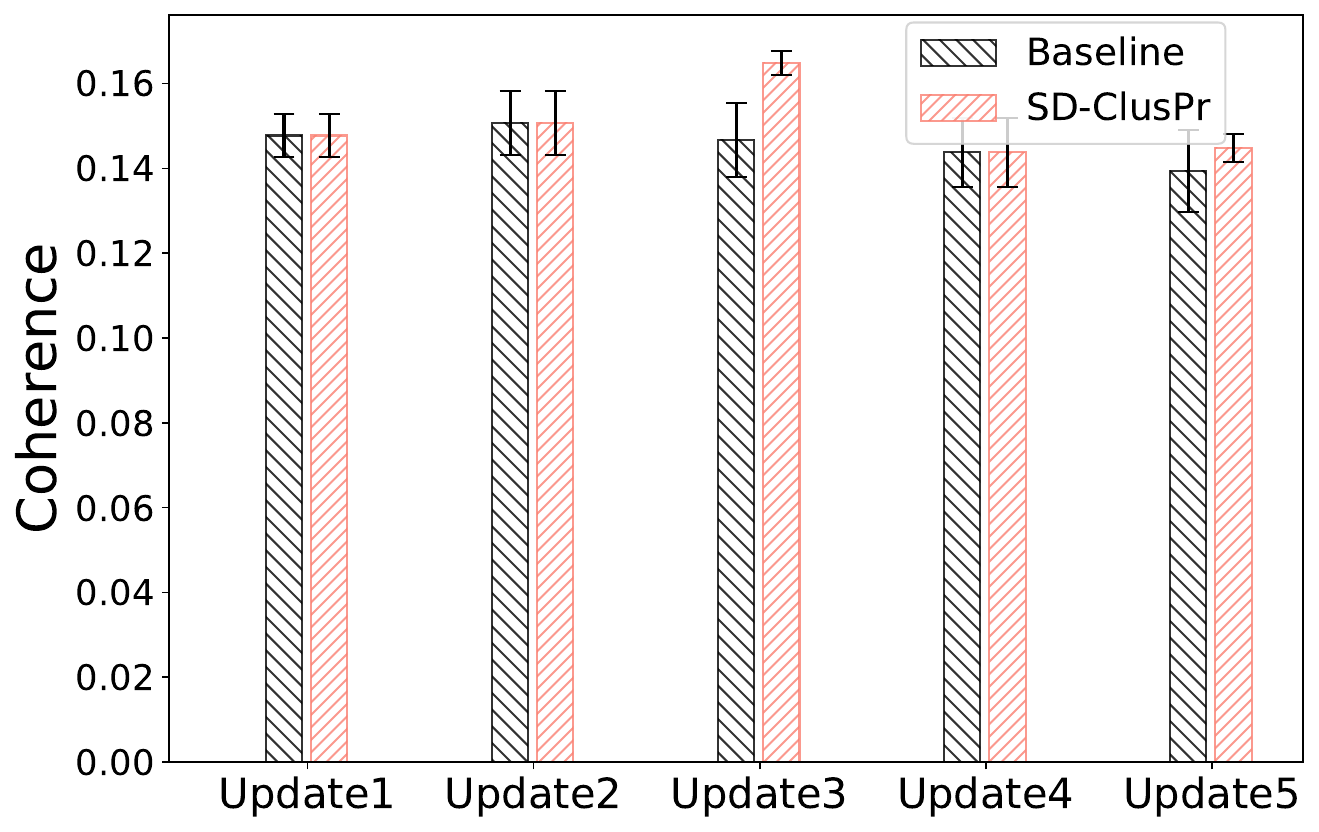}
\caption{RFC Dataset}
\label{fig: semidyna_rfc}
\end{subfigure}\hfill
\begin{subfigure}{.33\textwidth}
\centering
\includegraphics[width=\linewidth]{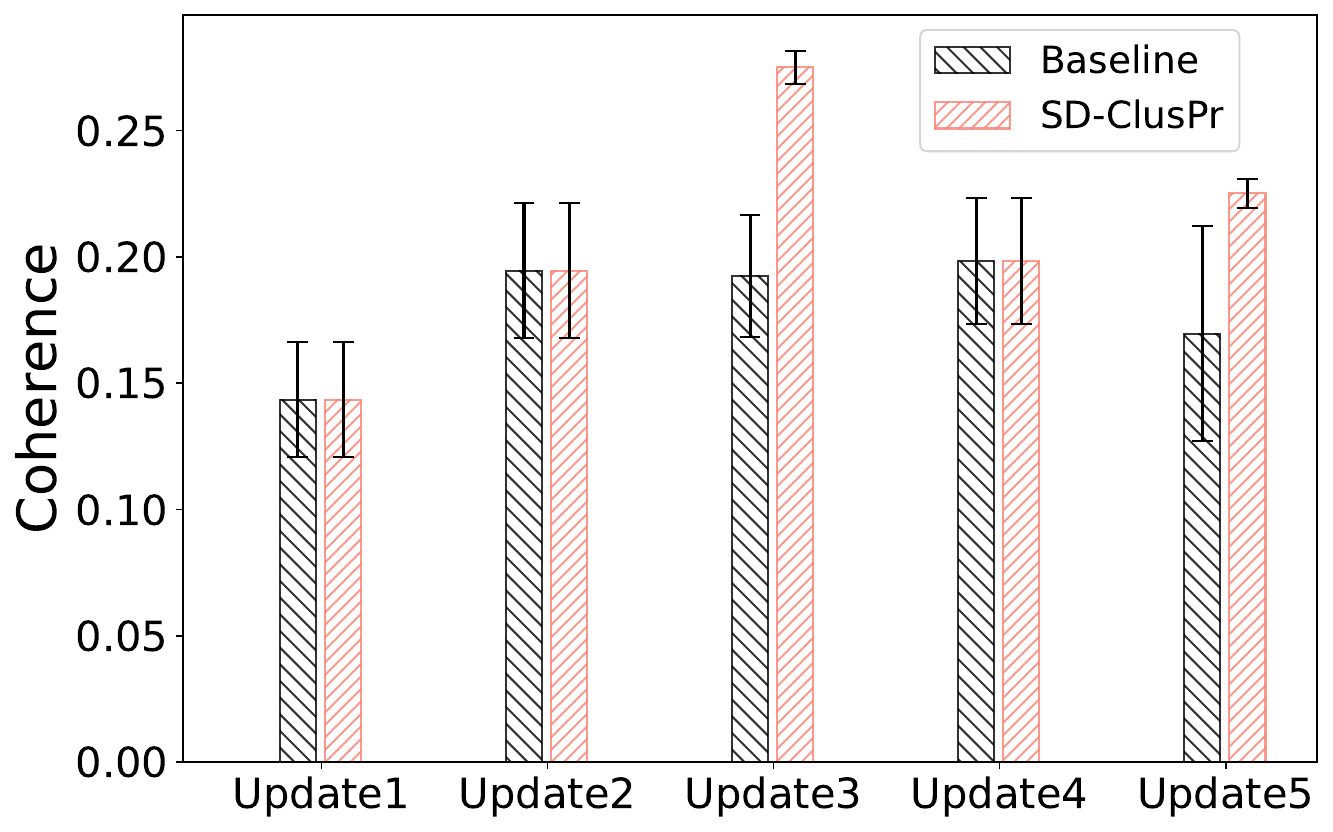}
\caption{ACCC Dataset}
\label{fig: semidyna_accc}
\end{subfigure}
\caption{Clusters' coherency for different updates of the three studied datasets when SD-\name~is applied with and without re-clustering option.}
\end{figure*}

\subsubsection{Evaluation of Clustering Coherency for Dynamic Schemes}
In this part, we analyze the effectiveness of dynamic clustering schemes (SD-\name~and FD-\name). We mention in Section~\ref{subsec:fdcs} that FD-\name~is a specific case of SD-\name. Hence, we only consider the SD-\name~scheme for evaluation. To this end, we leverage the three studied datasets and build subsets that each one serves as a batch update. Specifically, we consider an existing set of clusters based on $500$ documents for each dataset. Then, we sample five times to create a list of five updates that each one includes a set of documents. List $U$ includes the pairs of update names and the size of each update as follows: $U=<(U_1,25), (U_2,50), (U_3,100), (U_4,20), (U_5,200)>$. To assure that the results are not biased to any particular sample, we performed the sampling procedure 10 independent times and report the mean and 95\% confidence interval of the analysis in the results. The reason we designated $U_3$ and $U_5$ to be larger is to examine SD-\name~decision in re-clustering. To evaluate the scheme in terms of the cluster coherency, we build a \emph{baseline} version from SD-\name~that does not consider re-clustering. The baseline only performs clustering based on existing clusters (as explained in Algorithm~\ref{alg:semi}) to accommodate the new updates.

Figures~\ref{fig: semidyna_bbc},~\ref{fig: semidyna_rfc}, and~\ref{fig: semidyna_accc}, respectively, present cluster coherency of five different batch updates of \texttt{BBC},~\texttt{RFC}, and~\texttt{ACCC} respectively applying SD-\name~scheme. In Figure~\ref{fig: semidyna_bbc}, we observe that the coherency of clusters are decreased in baseline for $U_3$ whereas the coherency obtained for SD-\name~beats the previous by around 105\%. We observe the similar pattern of coherency variation for $U_5$. For baseline, the lowest coherency is obtained in $U_5$. On the contrary, in SD-\name, we observe around $115\%$ improvement in coherency for $U_5$.   

According to Figure~\ref{fig: main}, clusters formed for the \texttt{RFC} dataset shows the highest coherency. Similarly, in~\ref{fig: semidyna_rfc}, we observe the highest coherency for all updates in compare with other datasets. With respect to baseline, we observe that SD-\name~causes minor improvements in coherency of both $U_3$ and $U_5$. Since the documents are more domain-specific, clusters do not lose coherency significantly from one update to the other. As such, we do not observe significant improvements by SD-\name. 
Similar to \texttt{BBC} and \texttt{RFC}, in Figure~\ref{fig: semidyna_accc}, we observe improvement in the coherency for \texttt{ACCC} dataset. In particular, the improvement in coherency for $U_3$ and $U_5$ is approximately $45\%$ and $35\%$, respectively. 

From these experiments, we conclude that \name~scheme can improve the coherency of clustering even for dynamic datasets. Specifically, we observed that for sufficiently large batches, such as $U_3$ and $U_5$, SD-\name~decides to re-cluster that remarkably improves the clustering coherency. 



\section{CONCLUSIONS}\label{sec:conclsn}
In this research, we developed \name~as a solution for topic-based clustering of both static and dynamic unstructured encrypted big datasets. \name~approximates the number of clusters for a dataset within a feasible time complexity. For that purpose, \name~leverages the tokens' co-occurrences to measures the tendency of each token to stay with or segregate from other tokens and use that to estimate
the number of clusters. Next, we develop a probabilistic approach
to determine the center of each cluster and disseminate encrypted tokens to the most topically related cluster.
Experimental evaluations reveal that our clustering scheme for static datasets (S-\name) can improve the clustering coherency on average by $65\%$. Similarly, the scheme for semi-dynamic and dynamic datasets (SD-\name) can improve the coherency by $55\%$. By incorporating \name~within the context of a secure semantic search system, we learned that the more coherent and accurate topic-based clustering can improve the relevancy of search results. Although \name~outperforms other encryption-based clustering solutions, it is difficult for it to beat the schemes that operates on plain-text tokens. 
There are several avenues to extend this research work. 
One avenue is to identify the user search interest and leverage that to create more representative abstracts. Another avenue is to improve the cluster pruning component to reduce the search time operation.

\bibliographystyle{plain}
\bibliography{reference}


\end{document}